\documentclass[twocolumn,aps,amssymb,noshowpacs,superscriptaddress,nofootinbib]{revtex4}

\usepackage{amsmath,amssymb}
\usepackage{graphicx}
\usepackage{verbatim}
\usepackage{amsfonts}
\usepackage{dcolumn}
\usepackage{bm}
\usepackage[section]{placeins}
\usepackage{color}
\usepackage{mathrsfs}
\usepackage[colorlinks,bookmarks=true,citecolor=blue,linkcolor=red,urlcolor=blue]{hyperref}
\usepackage{afterpage}
\usepackage{ulem}
\usepackage{url} 

\newcommand{\bs}[1]{\boldsymbol{#1}}

\newcommand{\beq}{\begin{eqnarray} }
\newcommand{\eeq}{\end{eqnarray} }
\newcommand{\Beq}{\begin{eqnarray*} }
\newcommand{\Eeq}{\end{eqnarray*} }

\newcommand{\RNum}[1]{\uppercase\expandafter{\romannumeral #1\relax}}

\begin{document}
\draft

\title{Chiral spin liquid phase in an optical lattice at mean-field level}

\author{Jian Yang}
\affiliation{International Center for Quantum Materials, School of Physics, Peking
University, Beijing, 100871, China}
\affiliation{Beijing National Laboratory for Condensed Matter Physics and Institute of Physics, Chinese Academy of Sciences, Beijing 100190, China}

\author{Xiong-Jun Liu}
\email{xiongjunliu@pku.edu.cn}
\affiliation{International Center for Quantum Materials, School of Physics, Peking
University, Beijing, 100871, China}
\affiliation{Hefei National Laboratory, Hefei 230088, China}

\begin{abstract}
We study an optical Raman square lattice with $\mathrm{U}(1)$ synthetic gauge flux to show chiral spin liquid (CSL) phase for cold atoms
based on slave-rotor theory and spinon mean-field theory, respectively. An effective U($1$) gauge flux generated by Raman potentials plays a major role in realizing the CSL phase. By using slave-rotor techniques we find CSL phase at strong on-site Fermi Hubbard interacting regime. For large interacting regime we derive an effective spin model
including up to the four spin interactions. By spinon mean-field analysis it is shown that CSL phase is stabilized in the case of strong magnetic frustration. The two mean-field approximation methods give consistent phase diagrams and provide qualitative numerical evidence of the CSL phase.

\end{abstract}

\date{\today}
\maketitle

\section{Introduction}\label{sec1}
More than thirty years ago, Kalmeyer and Laughlin pointed out that the ground state wave function of the antiferromagnetic Heisenberg Hamiltonian in two-dimensional triangular lattice is equivalent to a fractional quantum Hall state for bosons~\cite{Kalmeyer}. The elementary excitations of the ground state
are neutral spin-$\frac{1}{2}$ particles. They obey
fractional (braiding) statistics and are called anyons.
Such fractional statistics is obtained in the system
where the parity and time-reversal symmetry are broken.
Motivated by this argument, chiral spin liquid (CSL) state was predicted at the end of 1980s.
In particular, for a Heisenberg spin Hamiltonian with both nearest neighbour and next nearest neighbour (diagonal) hopping in a two-dimensional square lattice, the CSL state is obtained and associated with the emergence of spin chirality interaction term $\mathbf{S}_{i}\cdot(\mathbf{S}_{j}\times\mathbf{S}_{k})$, with which the parity and time-reversal symmetries are violated~\cite{Wen1989}. 
Actually, the hopping through a minimum triangular loop composed of the three lattice sites ($i,j,k$) leads to a $\theta$ flux phase, which ensures the appearance of the above spin chirality interaction term. Then the spin degrees of freedom will experience an effective gauge field and quantum Hall effect occurs. In this sense, the CSL state is regarded as quantum Hall effect of spin degrees of freedom, whose boundary excitation energy bands are chiral gapless~\cite{Wen1991}. The topological structures of CSL are embodied in quantized Hall conductivity, a typical topological invariant.
Furthermore, the
CSL state can also be induced by the Dzyaloshinskii-Moriya (DM) interaction arising from spin-orbit coupling
\cite{diego2023prl}.
Due to the nontrivial topology of CSL,
much attention has been paid to theoretical explorations of the exotic topological phase in strong correlated systems \cite{Schro2007,Yao2007,Her2009,nabelqaf2009,SPK2011,Mess2012,lslcft2012,NY2013,parhamnabelcsl2014,Tigran2015,
Hickey2015,Chen2016,Hickey2016,xiongjun-16njp,Mess2017,Sopheak2018,HoiYinHui2019,
RJM2020,ashvin2021prb,chengang2021prr,qiu2021,zyh2021prl,
sch2022prb,kad2022prb,mer2022prr,song2023deconfined,
desr2023prb,bose2023prb,banerjee2023electromagnetic,diego2023prl}.

Among various attempts to detect novel topological phases in experiment, the platforms of ultra-cold atoms trapped in optical lattice
are very attractive since the optical lattice can simulate real crystalline structure with tunable lattice constant, potential barrier height, and onsite interaction that can be controlled by changing the optical lattice depths or magnetic Feshbach resonance. The strong correlated systems in condensed matter physics can be studied via quantum simulation of strong correlated states with ultra-cold atoms~\cite{Bloch2008,Schweizer_2019,Kohlert2019,Vijayan2020,Koepsell2021,Scherg_2021,Aidelsburger_2021,David2022,
Sompet_2022,bloch2022superfluid,Hirthe_2023}.

In recent years, the optical Raman lattice schemes\cite{Zhanglong_2018,Zhangdanwei_2018} have been proposed in theory and widely studied in experiment to generate synthetic gauge fields for cold atoms such that some novel topological phases can be detected. One approach is to adopt Raman couplings to create spin-orbit (SO) interactions ~\cite{xiongjun2014,WuSci2016,Huang2016,Meng2016,bz2018pra,Song2019np,xjl2020prl1,yhl2020scib,zyw2021scic,
wangbeibei2021,Ziegler_2022,Fulgado_Claudio_2023,xjl2023prl1}
of various types for ultra-cold atoms. 
Another is to use optical Raman lattice without spin-flip hopping between nearest and next nearest neighbour sites. When hopping along a closed path in the lattice, the accumulated non-trivial phase is equivalent to an effective
Aharonov-Bohm phase~\cite{Bloch2011,Bloch2013,Ketterle2013a,Ketterle2013b,Struck2013,HoiYinHui2019,ns2018prb}.
In comparison with the spin-flipped optical Raman lattices, the latter scheme can be achieved with far-off-resonant light, without suffering the
spontaneous emission due to near resonant lights.
The CSL phase was shown in the experimental setup
with $\mathrm{U}(1)$ synthetic gauge flux~\cite{xiongjun-16njp}.
A double-well square lattice and periodic Raman couplings can be generated by two incident plane-wave laser beams. The nearest neighbour spin-conserved hopping creates a nonzero phase. In
the single particle regime this model realizes a quantum
anomalous Hall (QAH) insulator (Chern insulator)~\cite{Haldane1988}
with a large gap-bandwidth ratio in the bulk and chiral gapless states
in the edge~\cite{xiongjun-10pra}. While for large Fermi Hubbard interactions it achieves an effective spin model containing spin chirality interaction term $\mathbf{S}_{i}\cdot(\mathbf{S}_{j}\times\mathbf{S}_{k})$. At mean-field approximation, the CSL phase appears by tuning parameters.

Slave-rotor formalism is a consistent framework to study correlated Fermi systems at strong interactions. The essence of this formalism is to interpret the physical variable associated with Mott transition as a quantum slave-rotor field dual to the local charge. The Mott insulator phase transition has been studied by applying the slave-rotor approach in correlated electron systems~\cite{Lee2005,StephanK-10prb,SPK2011,Hickey2015,SAA2019,TND2019,SPAA2020,HHL2020,PALee2021,dalalprr2021,wyhprb2022,
manuelprb2022,skim2023,song2023mottprb}.
In particular, the CSL phase was found at strong Hubbard interactions in honeycomb
lattice~\cite{SPK2011,Sopheak2018} via the slave-rotor approach. Moreover, this approach has been used to determine CSL phase in the platform of fermionic alkaline-earth-metal atoms trapped in an optical square lattice at $\mathrm{SU}(N)$ Hubbard interactions~\cite{Chen2016}.

In this paper, we systematically study the CSL phase in an improved
optical Raman square lattice with $\mathrm{U}(1)$ synthetic gauge flux
based on a scheme introduced in~\cite{bz2018pra}.
Compared with original optical Raman lattice setup~\cite{xiongjun-16njp} which
has a triangular loop to ensure generation of standing waves for the in-plane and
out-of-plane polarization components,
the improved setup has greater advantages on controllability and stability,
and can be well suitable for red- and blue-detuned optical lattices,
so it is of high feasibility in experiment.
Firstly, we apply slave-rotor mean-field approach for the Fermi Hubbard model and determine the existence of CSL. Charge and spin degrees of freedom are separated in this coupling regime, where charge degrees of freedom are in the Mott insulator state, but spin degrees of freedom form QAH state without long-range spin order, implying a CSL phase obtained.

On the other hand, for large Hubbard interactions, in addition to the third order correction term (spin chirality interaction term) in the effective spin model, we investigate the effect of fourth order couplings (four spin interaction terms). Among all the four spin interaction terms, an important term with four spins located at the four lattice sites of a minimum square plaquette are expected to have qualitative effect on the CSL phase. Because of the $\pi$ flux phase through hopping around the minimum square plaquette, the interesting results are obtained. We show that more consistent CSL phase diagram is obtained when relevant four-spin interactions are taken into account in the effective spin model.

The paper is organized as follows. In section \ref{TBM}, we review some basic properties of improved optical Raman lattice
and introduce
the Fermi Hubbard interaction. It can be shown that there is a topological phase transition between a normal insulator and a QAH insulator by adjusting parameters in single-particle spectra.
In section \ref{srmf} the basic idea of slave-rotor theory~\cite{florens-02prb165111,florens-04prb035114,zhaopara,StephanK-10prb} is presented.
Then in sections \ref{qahcsl} and \ref{cslmag} we apply the slave-rotor approach to solve the self-consistent equations at the boundary of CSL phase. Then the global phase diagram of CSL is shown in section \ref{mfphase}. At the same time we study the CSL phase in effective spin model by spinon mean-field calculation in section \ref{spinmod}.
Section \ref{sec:sum} is devoted to the conclusion and discussion.
Throughout the paper we consider half-filled case at zero temperature ($T=0$).


\section{Tight-binding model and general considerations}\label{TBM}

\begin{figure}[ht]
\includegraphics[width=1\columnwidth]{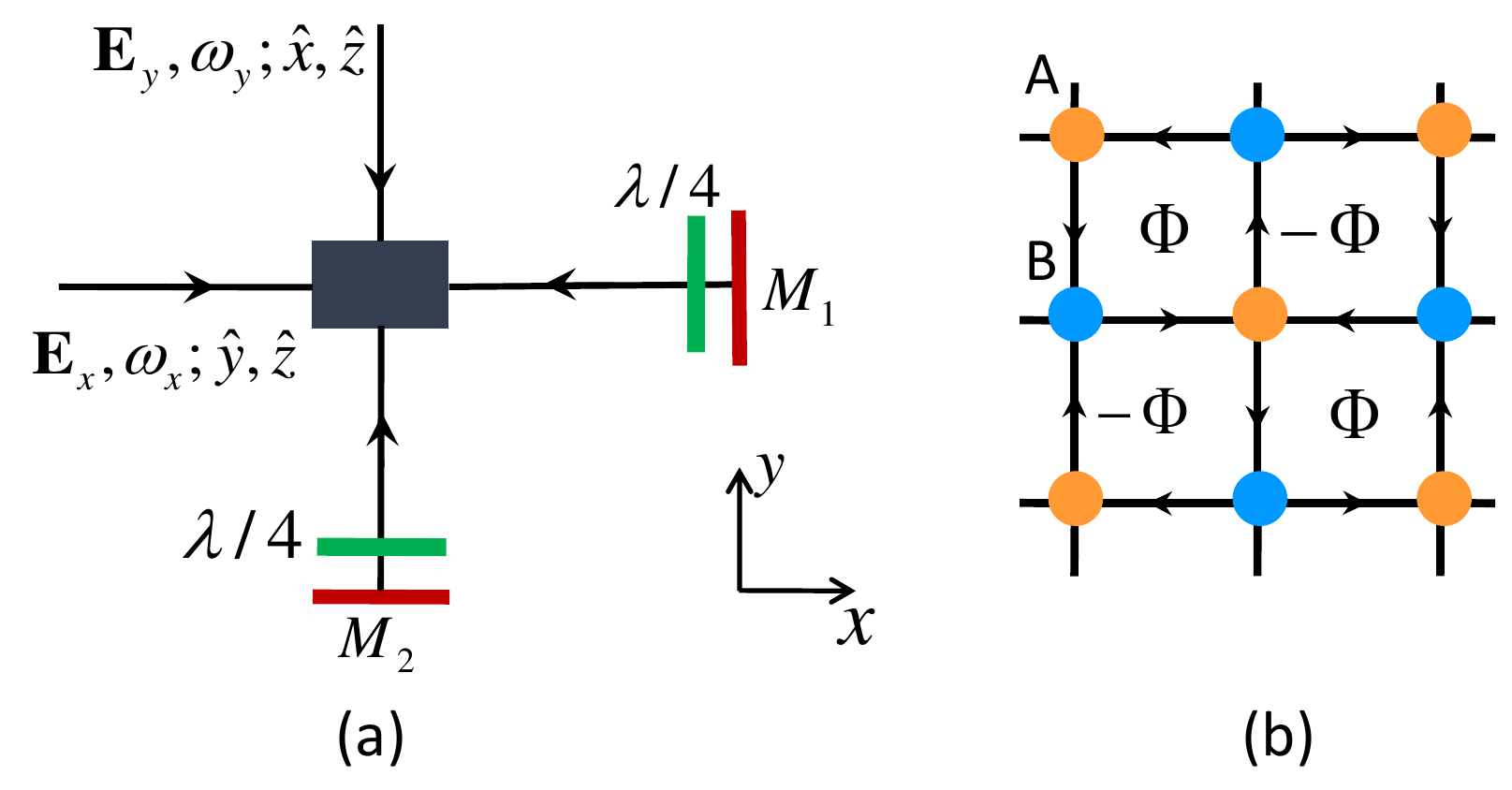}
\caption{(Color online)
(a)An incident plane-wave laser beam propagating along $x$ direction
with frequency $\omega_{x}$
has nonzero $\hat{y}$ and $\hat{z}$ linear polarization components,
and an incident plane-wave laser beam propagating along $y$ direction
with frequency $\omega_{y}$
has nonzero $\hat{x}$ and $\hat{z}$ linear polarization components.
With the two $\frac{1}{4}$-wave plates and reflections by two mirrors $M_{1,2}$,
two Raman potentials can be generated.
(b)The optical Raman square lattice has an energy offset between A and B
sublattices, and the nearest neighbour hopping caused by Raman potentials
$V_{R_{x}}$ and $V_{R_{y}}$.
The Raman transitions also generate a staggered
flux pattern for the nearest-neighbour hopping.}
\label{lattice}
\end{figure}

We begin with the anisotropic two-dimensional ($2$D) optical square lattice in Fig.\ref{lattice} which is realized by
the experimental setup depicted in Ref.~\cite{bz2018pra}.
Here
an incident plane-wave laser beam $\mathbf{E}_{x}$ from $x$ direction
with frequency $\omega_{x}$
has nonzero $\hat{y}$ and $\hat{z}$ linearly polarized components,
while an incident plane-wave laser beam $\mathbf{E}_{y}$ from $y$ direction
with frequency $\omega_{y}$
has nonzero $\hat{x}$ and $\hat{z}$ linearly polarized components.
Two $\frac{1}{4}$-wave plates are placed in the path from mirror $M_{1}$ to lattice center
and the path from mirror $M_{2}$ to lattice center, respectively,
in Fig.\ref{lattice}(a), which lead to
additional $\frac{\pi}{2}$-phase shift for the $\hat{z}$ polarized component.
$\mathbf{E}_{x}$ and $\mathbf{E}_{y}$ generate the standing waves as
\beq
\mathbf{E}_{x}&=&\hat{y}E_{xy}e^{i\phi_{xy}}\cos(k_{0}x)
+i\hat{z}E_{xz}e^{i\phi_{xz}}\sin(k_{0}x),\label{}
\\
\mathbf{E}_{y}&=&\hat{x}E_{yx}e^{i\phi_{yx}}\cos(k_{0}y)
+i\hat{z}E_{yz}e^{i\phi_{yz}}\sin(k_{0}y),\label{}
\eeq
where
$\phi_{xy}/\phi_{xz}$ is the initial phase of $E_{xy}/E_{xz}$
polarization component,
$\phi_{yx}/\phi_{yz}$ is the initial phase of
$E_{yx}/E_{yz}$ polarization component,
$k_{0}=\frac{\omega_{x}}{c}\approx\frac{\omega_{y}}{c}$,
$E_{\mu\nu}$ ($\mu,\nu=x,y,z$) is the amplitude of standing wave
from $\mu$ direction with $\nu$ polarized components.
The optical square lattice potential
can be formed:
\beq
V_{\rm sq}(x,y)=V_{0x}\cos^{2}(k_{0}x)+V_{0y}\cos^{2}(k_{0}y),
\eeq
where the amplitudes $V_{0x}\propto\frac{E_{xy}^{2}-E_{xz}^{2}}{\Delta}$ and
$V_{0y}\propto\frac{E_{yx}^{2}-E_{yz}^{2}}{\Delta}$
with red or blue detuning $\Delta$.

Two Raman couplings are also induced by $\mathbf{E}_{x}$ and $\mathbf{E}_{y}$,
and the Raman potentials take the forms~\cite{bz2018pra}
\beq
V_{R_{x}}&=&V_{R_{10}}\cos(k_{0}x)e^{-i\phi_{xy}}\sin(k_{0}y)e^{i\phi_{yz}},
\label{rmpx}\\
V_{R_{y}}&=&V_{R_{20}}\sin(k_{0}x)e^{-i\phi_{xz}}\cos(k_{0}y)e^{i\phi_{yx}},
\label{rmpy}
\eeq
with the amplitudes
\beq
V_{R_{10}}\propto E_{xy}E_{yz},
V_{R_{20}}\propto E_{yx}E_{xz}.
\eeq
Thus, the anisotropic $2$D optical Raman square lattice
can be realized with
a staggered energy offset between $A$ and $B$ sublattices
in Fig.\ref{lattice}(b).
Since $E_{xy}$ and $E_{xz}$ come from the same incident laser beam,
we can set that $\phi_{xy}=\phi_{xz}\equiv\phi_{y}$.
Similarly, $\phi_{yx}=\phi_{yz}\equiv\phi_{x}$.
As can be seen below that a finite magnitude of $\phi_{x}-\phi_{y}$
controllable in experiment, results in a nonzero staggered flux pattern
for the square lattice in Fig.\ref{lattice}(b).

Here we consider only the $s$-orbital wave functions $\psi^{(\vec j)}_{\mu,s}(\bold r)$ at sublattice $A$ ($\mu=a$) and $B$ ($\mu=b$), which are of even parity.
From
Raman potentials $V_{Rx,Ry}$ in Eqs.\eqref{rmpx} and \eqref{rmpy},
the Raman potentials are parity odd relative to each
lattice-site center.
Due to these symmetry properties,
for the $s$-orbital bands,

(i) The Raman potentials $V_{Rx,Ry}$ can induce the hopping
between nearest neighbour $A$ and $B$ sublattices, but can not induce next
nearest neighbour ($AA/BB$) hopping.

(ii) The Raman potential $V_{Rx}$ ($V_{Ry}$)
leads to the hopping along $x$ ($y$) direction, which is accompanied by a phase $\phi_{y/x}$ ($-\phi_{y/x}$), when the hopping is toward (away from) $B$ sites.
In experiment, we can readily set that $\phi_x-\phi_y=2\phi_0$,
which is equivalent to $\phi_x=-\phi_y=\phi_0$,
so the hopping along a closed paths described by arrows in Fig.\ref{lattice}(b) acquires a phase $4\phi_0$,
which is equivalent to an effective Aharonov-Bohm phase.
This leads to a uniform $\mathrm{U}(1)$ gauge field with magnetic flux through each plaquette being $4\phi_0$
but with alternating sign along $x$ and $y$ axis, thus
a staggered flux configuration with the flux $|\Phi|=4\phi_0$
in each square plaquette~\cite{Bloch2011}.

(iii) Owing to the odd parity of Raman potentials $V_{Rx,Ry}$, the hopping from one site to its leftward (upward)
neighboring site has an additional minus sign relative to the hopping to its rightward (downward) neighboring site.

\begin{figure}[ht]
\includegraphics[width=0.6\columnwidth]{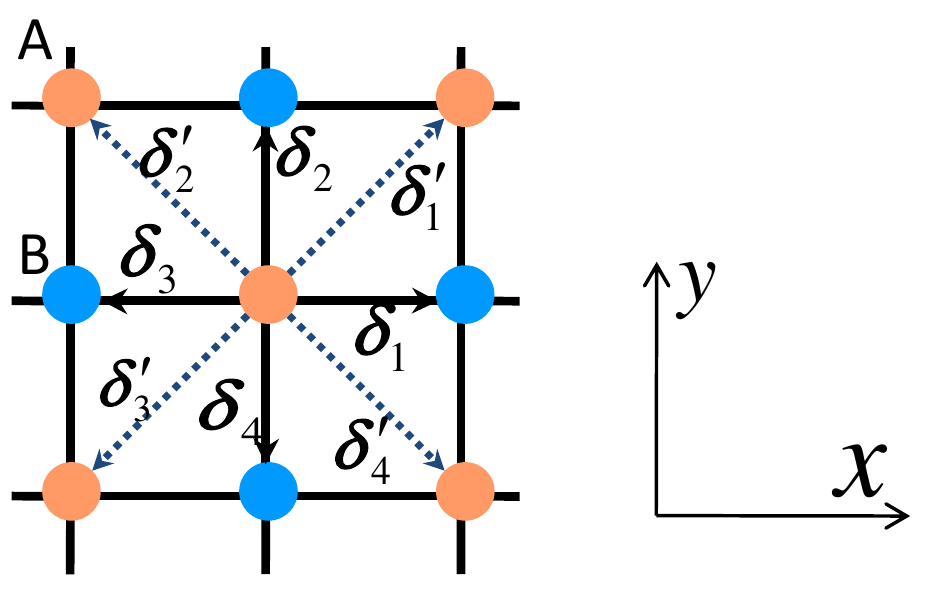}
\caption{(Color online) 2D optical Raman lattice divided into two sublattices A and B.
The four nearest neighbour vectors $\bs{\delta_{i}} (i=1,2,3,4)$ are connecting the two sublattices (solid arrows),
and the four next nearest neighbour vectors $\bs{\delta'_{i}} (i=1,2,3,4)$ are also shown (dashed arrows).}
\label{sqlattice}
\end{figure}

With the above properties, we can obtain the $s$-band tight-binding Hamiltonian
\begin{eqnarray}\label{tbHam1}
H&=&-\sum_{\langle i j\rangle}(t_{ i j}e^{i\phi_{ i  j}} \hat{c}_{b, i}^\dag \hat{c}_{a, j}+{\rm H.c.})
\nonumber\\
&&
-\sum_{\langle\langle i j\rangle\rangle}\sum_{\mu=a,b}t'_{\mu, i j} \hat{c}_{\mu, i}^\dag \hat{c}_{\mu, j}
\nonumber\\
&&
+m_z\sum_{ i}(\hat{c}^\dag_{a, i} \hat{c}_{a, i}-\hat{c}^\dag_{b, i} \hat{c}_{b, i}).
\end{eqnarray}
Here
$ \hat{c}_{\mu, i}$ is the fermionic annihilation operator on sublattice $A$ (for $\mu=a$) and $B$ (for $\mu=b$).
The nearest neighbour vectors $\bs{\delta}_{1}(-\bs{\delta}_{3})=(a,0)$, $\bs{\delta}_{2}(-\bs{\delta}_{4})=(0,a)$
and the next nearest neighbour vectors $\bs{\delta}_{1}'(-\bs{\delta}_{3}')=(a,a)$, $\bs{\delta}_{2}'(-\bs{\delta}_{4}')=(-a,a)$
are shown in Fig.\ref{sqlattice}, with $a$ the lattice constant.
$\langle i j\rangle$ and $\langle\langle i j\rangle\rangle$
denote nearest neighbour and next nearest neighbour sites, respectively.
The nearest neighbor and diagonal hopping coefficients
(excluding the hopping phases),
$t_{ ij}$ and $t'_{\mu, i j}$ can been calculated like in
Refs.\cite{xiongjun-16njp,xiongjun2014,bz2018pra}:
\beq
t_{i,i\pm 1_{x}}&=&V_{R_{10}}
\int d^{2}\mathbf{r}\psi_{b,s}^{(i_{x},i_{y})}(\mathbf{r})
\cdot
\nonumber\\
&&
\cos(k_{0}x)\sin(k_{0}y)
\psi_{a,s}^{(i_{x}\pm 1,i_{y})}(\mathbf{r}),
\\
t_{i,i\pm 1_{y}}&=&V_{R_{20}}
\int d^{2}\mathbf{r}\psi_{b,s}^{(i_{x},i_{y})}(\mathbf{r})
\cdot
\nonumber\\
&&
\cos(k_{0}y)\sin(k_{0}x)
\psi_{a,s}^{(i_{x},i_{y}\pm 1)}(\mathbf{r}),
\\
t'_{\mu, ij}&=&
\int d^{2}\mathbf{r}\psi_{\mu,s}^{(i_{x},i_{y})}(\mathbf{r})
[\frac{\mathbf{p}^{2}}{2m}+V_{\mathrm{sq}}(\mathbf{r})]
\psi_{\mu,s}^{(i_{x}\pm 1,i_{y}\pm 1)}(\mathbf{r}),
\nonumber\\
\eeq
where $\bold p$ is momentum of atom and $m$ is atom mass.
It can be verified that
\beq
t_{ i, i\pm1_x}=\pm(-1)^{i_x}t_0,
t_{ i, i\pm1_y}=\mp(-1)^{i_x}t_0,
t'_{\mu, i  j}=t'_\mu,
\eeq
with
$t_0=V_{R_{10}}\int d^2\bold r\psi_{b,s}^{(0,0)}(\bold r)\cos(k_0x)\sin(k_0y)\psi_{a,s}^{(1,0)}(\bold r)$ and
$t'_\mu=\int d^2\bold r\psi_{\mu,s}^{(0,0)}(\bold r)\bigr[\frac{\bold p^2}{2m}+V_{\rm sq}(\bold r)\bigr]\psi_{\mu,s}^{(1,1)}(\bold r)$ ($\mu=a,b$).
The staggered sign $(-1)^{i_x}$ can be absorbed by transforming sublattice $B$ annihilation operator $\hat{c}_{b, j}$ into $e^{i\pi \frac{x_j}{a}}\hat{c}_{b, j}$.
In terms of the new operator, the diagonal hopping coefficient $t'_b$ acquires an additional minus sign $-t'_b$.
The hopping phase $\phi_{ij}=\nu_{ij}\phi_0$ with $\nu_{ij}=1$ ($-1$) for hopping along (opposite to) the marked direction in Fig.\ref{lattice}(b), and $m_z$ is the Zeeman term.
Since $\psi_{a,s}$ and $\psi_{b,s}$ have the same spatial distribution, we may set $t'_a\approx t'_b$.

In order to diagonalize the Hamiltonian (\ref{tbHam1}) we transform it into momentum space via
\begin{equation}\label{fouriertrans}
\hat{c}_{a,b}(\bold r_{i})=\frac{1}{\sqrt{N}}\sum_{\bold k}e^{i \bold k \cdot \bold r_{i}}\hat{c}_{a,b}(\bold k),
\end{equation}
here $N$ is the number of unit cells, the sum over $\bold{k}$ is on the first Brillouin zone (FBZ). We get Hamiltonian in matrix form $H=\sum_{\bold k}\hat {\mathcal C}^{\dag}(\bold
k)\mathcal{H}(\bold k)\hat {\mathcal C}(\bold k)$ with $\hat
{\mathcal C}(\bold k)=(\hat c_{a}(\bold k), \hat c_{b}(\bold k))^T$ and
\begin{eqnarray}\label{BlochHamiltonian1}
{\cal H}(\bold k)=d_x(\bold k)\sigma_x+d_y(\bold k)\sigma_y+d_z(\bold k)\sigma_z,
\end{eqnarray}
with the coefficients
\begin{eqnarray}
d_x&=&-2t_0\sin\phi_0(\sin k_xa+\sin k_ya), \label{dxcoeff}\\
d_y&=&-2t_0\cos\phi_0(\sin k_xa-\sin k_ya), \label{dycoeff}\\
d_z&=&m_z-2(t'_a+t'_b)\cos( k_xa)\cos( k_ya),\label{dzcoeff}
\end{eqnarray}
where $\sigma_{x,y,z}$ are three Pauli matrices for sublattices.
It is clear that as long as $\phi_0\neq n\pi$ with $n\in\mathbb{Z}$, the time-reversal symmetry of system is broken,
thus it can give rise to QAH effect.

Based on the band structure one can manifest QAH effect.
The Hamiltonian has two energy bands given by $\mp\varepsilon=\mp\sqrt{d_x^2+d_y^2+d_z^2}$. Their corresponding eigenstates
are (up to an arbitrary phase):
\begin{eqnarray}
|u_-\rangle&=&\left(\begin{array}{c}\alpha_{-}\\[5pt]
\beta_{-}\end{array}\right)
=\left(\begin{array}{c}\sin\frac{\theta}{2}e^{-i\varphi}\\[5pt]
-\cos\frac{\theta}{2}\end{array}\right),\label{eigsta-}\\
|u_+\rangle&=&\left(\begin{array}{c}\alpha_{+}\\[5pt]
\beta_{+}\end{array}\right)
=\left(\begin{array}{c}\cos\frac{\theta}{2}e^{-i\varphi}\\[5pt]
\sin\frac{\theta}{2}\end{array}\right),\label{eigsta+}
\end{eqnarray}
where the mixing angles are defined by $\tan\theta=\frac{\sqrt{d_x^2+d_y^2}}{d_z}$,
$\tan\varphi=\frac{d_y}{d_x}$. Now the Hamiltonian (\ref{tbHam1}) can be diagonalized:
\begin{eqnarray}\label{diaBlochHamiltonian}
H&=&\sum_{\bold k}\big(\hat c^{\dagger}_{a}(\bold k), \hat c^{\dagger}_{b}(\bold k)\big)
\mathcal{H}(\bold k)
\left(\begin{array}{c}\hat c_{a}(\bold k)\\[5pt]
\hat c_{b}(\bold k)\end{array}\right)\nonumber\\
  &=&\sum_{\bold k}\big( \hat{l}_{\bold k}^\dagger, \hat{u}_{\bold k}^\dagger\big)
\left(\begin{array}{cc}-\varepsilon&0\\[5pt]0&\varepsilon\end{array}\right)\left(\begin{array}{c}\hat{l}_{\bold k}\\[5pt]
\hat{u}_{\bold k}\end{array}\right),
\end{eqnarray}
here the new set of operators $\hat{l}_{\bold k}$
and $\hat{u}_{\bold k}$ associated with lower and upper bands are introduced.

To evaluate the gap, we find if $m_z$ is tuned from $m_z>2(t'_a+t'_b)$ to $m_z=2(t'_a+t'_b)$, the band gap is closed at Dirac point $\bold k=(0,0)$, and further tuning it from $m_z=2(t'_a+t'_b)$ to $m_z<2(t'_a+t'_b)$, the gap is reopened. Similarly, adjusting $m_z$ from $m_z<-2(t'_a+t'_b)$ to $m_z>-2(t'_a+t'_b)$, the gap is closed and then reopened at Dirac point $\bold k=(0,\frac{\pi}{a})$.

If the Fermi energy (chemical potential) lies in the gap, only the lower band is occupied, we may use the eigenstate $|u_-\rangle$ to calculate the anomalous Hall conductivity (AHC), which reads:
\begin{equation}\label{AHC1}
\sigma^{H}_{xy}
=\frac{1}{h}\int_{{\rm FBZ}} dk_xdk_y\frac{\bold n\cdot(\partial_{k_x}\bold n\times\partial_{k_y}\bold n)}{4\pi}
=\frac{C_1}{h},
\end{equation}
where $\bold n=(d_x,d_y,d_z)/(d_x^2+d_y^2+d_z^2)^{1/2}$. $C_1$ is the first Chern number, a quantized
topological invariant~\cite{KOHMOTO1985343,NQ2010} defined on the FBZ.
By derivation one can show
\begin{eqnarray}\label{chern1}
C_1 &=&+1 , \text{for  $0<\phi_0<\frac{\pi}{2}, |m_z|<2(t'_a+t'_b)$,}\nonumber\\
C_1 &=&-1 , \text{for $\frac{\pi}{2}<\phi_0<\pi, |m_z|<2(t'_a+t'_b)$,}\nonumber\\
C_1 &=&0 ,  ~~\text{for $\phi_0=\frac{n\pi}{2}(n\in\mathbb{Z})$ or $|m_z|>2(t'_a+t'_b)$.}\nonumber\\
\end{eqnarray}
It indicates that topological phase (QAH insulator) exists only when $|m_z|<2(t'_a+t'_b)$ and $\phi_0\neq\frac{n\pi}{2}(n\in\mathbb{Z})$.
which has gapped bulk states \cite{xiongjun-16njp} but supports gapless edge states on the boundaries of the system \cite{xiongjun-10pra}.

Next we will take into account a spin-$\frac{1}{2}$ two-copy version of QAH model. The two copies can be obtained
from two subspaces $\Gamma_1$ and $\Gamma_2$ of a system. $\Gamma_1$ subspace is described by a Hamiltonian
corresponding to spin-up state:
\begin{equation}\label{spinupblochH}
H_{\Gamma_1}=\sum_{\bold k}\big(\hat c^{\dagger}_{a\uparrow}(\bold k), \hat c^{\dagger}_{b\uparrow}(\bold k)\big)
\mathcal{H}(\bold k)
\left(\begin{array}{c}\hat c_{a\uparrow}(\bold k)\\[5pt]
\hat c_{b\uparrow}(\bold k)\end{array}\right),
\end{equation}
and $\Gamma_2$ subspace is described by a Hamiltonian
corresponding to spin-down state:
\begin{equation}\label{spindownblochH}
H_{\Gamma_2}=\sum_{\bold k}\big(\hat c^{\dagger}_{a\downarrow}(\bold k), \hat c^{\dagger}_{b\downarrow}(\bold k)\big)
\mathcal{H}(\bold k)
\left(\begin{array}{c}\hat c_{a\downarrow}(\bold k)\\[5pt]
\hat c_{b\downarrow}(\bold k)\end{array}\right),
\end{equation}
here in Eqs.\eqref{spinupblochH} and \eqref{spindownblochH}, $\mathcal{H}(\bold k)$ is given by Eq.\eqref{BlochHamiltonian1}.
When Fermi energy is inside the gap and both $\Gamma_1$ and $\Gamma_2$ subsystems are half filled,
each subspace forms a QAH insulator with the same Chern number. Since $\Gamma_1$ and $\Gamma_2$ subspaces decouple, we have the total Hamiltonian:
$H=\sum_{\bold k}\hat {\mathcal C}^{\dag}(\bold
k)\mathcal{H}(\bold k)\hat {\mathcal C}(\bold k)$ with $\hat
{\mathcal C}(\bold k)=(\hat c_{a\uparrow}(\bold k), \hat c_{b\uparrow}(\bold k),\hat c_{a\downarrow}(\bold k), \hat c_{b\downarrow}(\bold k))^T$ and
\begin{eqnarray}\label{BlochHamiltonian2}
\mathcal{H}(\bold k)&=&\sum_{\alpha=x,y,z}d_{\alpha}(\bold k)I\otimes\sigma_\alpha.
\end{eqnarray}

The main task of this paper is to investigate the effect of a repulsive Fermi Hubbard interaction on the above spin-$\frac{1}{2}$ two-copy version of QAH model. The Hubbard term can be written as:
\begin{equation}\label{hubbard1}
H_I = U \sum_i n_{i\uparrow} n_{i\downarrow}\ ,
\end{equation}
where $U$ is the strength of Hubbard interaction.
In the case of half filling we may rewrite the Hubbard interaction as
\begin{equation}\label{hubbard2}
H_I = \frac{U}{2} \sum_i \left( \sum_\sigma n_{i\sigma} - 1 \right)^2\ .
\end{equation}
This formulation often appears in the slave-rotor theory.
In what follows, we set lattice constant $a=1$, next nearest neighbour hopping coefficients $t'_a=t'_b=t_1$ and the Zeeman term $m_z=\omega_{x}-\omega_{y}=0$ for resonant Raman process.

\section{slave-rotor mean-field formalism}\label{srmf}

Now we give the general idea of slave-rotor approach.
According to Ref. \cite{florens-02prb165111,florens-04prb035114,zhaopara,StephanK-10prb}, the original fermion operator $\hat{c}_{i\sigma}$ will be rewritten by a product of a spin-$\frac{1}{2}$ spinon (auxiliary fermion) operator $\hat{f}_{i\sigma}$ and a charged rotor $e^{i\theta_i}$,
\begin{equation}\label{rewfermi}
\hat{c}_{i\sigma} = e^{i\theta_i}\hat{f}_{i\sigma}.
\end{equation}
Based on this representation, the phase variable $\theta_{i}$ is conjugate to the total charge.
In terms of the new variable, the quartic Hubbard interaction term \eqref{hubbard2} between the fermions is expressed by a simple kinetic term $\hat{L}_{i}^2$, where the angular momentum operator $\hat{L}_i\propto i\partial_{\theta_i}$ is a conjugate momentum of a quantum O(2) rotor field $\theta_i$.
State vectors in the new Hilbert space take the form $|\Psi\rangle=|\Psi_f\rangle|\Psi_\theta\rangle$. The new rotor-spinon Hilbert space is enlarged compared to the original one since there exist unphysical states. To eliminate these unphysical states we have to impose a constraint about operators,
\begin{equation}\label{constraint1}
\sum_\sigma \hat{f}_{i\sigma}^\dagger \hat{f}_{i\sigma}^{\phantom{\dagger}} + \hat{L}_i =1\ .
\end{equation}
The hopping terms of Hamiltonian are rewritten as
\begin{eqnarray}\label{tbHamMF}
H&=&-\sum_{\langle i j\rangle\sigma}t_{ i j}e^{i\phi_{ i j}} \hat{f}_{i\sigma}^\dag \hat{f}_{j\sigma}e^{-i\theta_{ij}}\nonumber\\
&&-\sum_{\langle\langle i j\rangle\rangle}\sum_{\sigma}t_1\hat{f}_{ i\sigma}^\dag \hat{f}_{ j\sigma}e^{-i\theta_{ij}},
\end{eqnarray}
where $\theta_{ij}\equiv \theta_i - \theta_j$.
Moreover, we replace the phase field $e^{i\theta_i}$ representing the O($2$) degree of freedom by a complex bosonic field $X_i(\tau)$ which is constrained by
\begin{equation}\label{constraint2}
|X_i(\tau)|^2=1,
\end{equation}
here the imaginary time $\tau=it$.

The above mean-field formalism can be used to treat strong correlated fermionic systems, where the bosonic field $X_i(\tau)$ is related to Mott transition.

\section{Transition from QAH state to CSL}\label{qahcsl}
In this section we use the slave-rotor mean-field formalism to determine the phase boundary between QAH state and CSL. In QAH state, the rotor is condensed and
the original fermion operator $\hat{c}_{i\sigma}$ is proportional to the spinon operator $\hat{f}_{i\sigma}$. In other words, the degrees of freedom of rotor and spinon are not separated. The whole system is a QAH insulator.
While in the CSL state , the charge degrees of freedom form a Mott insulator state, but the spinon is described by a Hamiltonian which is very similar to the Hamiltonian \eqref{BlochHamiltonian2} of the  spin-$\frac{1}{2}$ two-copy version of QAH model. In this respect the rotor undergoes a phase transition from superfluid to Mott insulator. Furthermore, no symmetry breaking orders (such as magnetic orders) appear in the CSL state.

The Hubbard interaction term \eqref{hubbard2} is rewritten in terms of the new variables $e^{i\theta_{i}}$ and $\hat{f}_{i\sigma}$ as
\begin{equation}\label{hubbard3}
H_I = \frac{U}{2} \sum_{i} \left( \sum_{\sigma} n_{i\sigma} - 1 \right)^{2}=\frac{U}{2}\sum_{i}\widehat{L}_{i}^{2},
\end{equation}
where we use the constraint \eqref{constraint1} and $n_{i\sigma}=\hat{f}^{\dagger}_{i\sigma}\hat{f}_{i\sigma}$.

The slave-rotor Hamiltonian reads
\begin{eqnarray}\label{}
H&=&-\sum_{\langle i j\rangle\sigma}t_{ i j}e^{i\phi_{ i  j}} \hat{f}_{i\sigma}^\dag \hat{f}_{j\sigma}e^{-i\theta_{ij}}\nonumber\\
&&-\sum_{\langle\langle i j\rangle\rangle}\sum_{\sigma}t_1\hat{f}_{ i\sigma}^\dag \hat{f}_{ j\sigma}e^{-i\theta_{ij}}\nonumber\\
&& -\mu\sum_{i,\sigma}\hat{f}^{\dagger}_{i\sigma}\hat{f}_{i\sigma}+\frac{U}{2}\sum_{i}\widehat{L}_{i}^{2}.
\end{eqnarray}
where $\mu$ is chemical potential.
Then the mean-field decomposed action is
\begin{eqnarray}
S_0&=&\int_0^\beta d\tau \Big[ \sum_{i\sigma} \hat{f}_{i\sigma}^\dagger \left( \partial_\tau - \mu + h_i \right)\hat{f}_{i\sigma}  + \sum_i \rho_{i} |X_i|^2\nonumber\\
& &+ \frac{1}{2U}\sum_i[(i\partial_\tau +i h_i)X_i^\ast][(-i\partial_\tau + ih_i)X_i]\nonumber\\
& &
+\sum_i \left( -h_i + \frac{h_i^2}{2U}\right)  +H_X +H_X'+H_f+H_f'\Big],\nonumber\\
\label{decomaction1}
\end{eqnarray}
here $\beta=\frac{1}{k_{B}T}=\frac{\tau}{\hbar}$. We have to fulfill the constraints \eqref{constraint1} and \eqref{constraint2}
with the Lagrange multipliers $h_i$ and $\rho_i$, respectively.
In calculation, \eqref{constraint1} and \eqref{constraint2} are treated on average \cite{florens-04prb035114,zhaopara,StephanK-10prb}, $\sum_\sigma
\langle
\hat{f}_{i\sigma}^\dagger \hat{f}_{i\sigma}^{\phantom{\dagger}}\rangle +
\langle
\hat{L}_i\rangle =1$,
$\langle|X_i(\tau)|^2\rangle=1$, so the Lagrange multipliers $h_i\equiv h$ and $\rho_i\equiv\rho$ are not local site-dependent.

Since it contains $A$, $B$ sublattices, the effective Hamiltonians for rotor and spinon parts are
\begin{eqnarray}
H_X&=&-Q_X\sum_{\langle i j\rangle}X_{i}^{b\ast}X_{j}^{a}+ {\rm c.c.},\label{Hx}\\
H_f&=&-Q_f\sum_{\langle i j\rangle\sigma}t_{ i j}e^{i\phi_{ i  j}} \hat{f}_{i\sigma}^{\dag} \hat{f}_{j\sigma}
+ {\rm H.c.} ,\label{Hf}\\
H_X'&=&-Q_X'\sum_{\langle\langle i j\rangle\rangle}X_{i}^{\ast}X_{j},\label{Hxprm}\\
H_f'&=&-Q_f'\sum_{\langle\langle i j\rangle\rangle}\sum_{\sigma}t_1\hat{f}_{ i\sigma}^\dag \hat{f}_{ j\sigma}\label{Hfprm}.
\end{eqnarray}
The mean-field parameters associated with the decomposition are given by
\begin{eqnarray}
\label{def:qx}
Q_X &=&\langle\sum_{\sigma}t_{ i j}e^{i\phi_{ i  j}}\hat{f}_{i\sigma}^{b\dag}\hat{f}_{j\sigma}^{a}\rangle , \\
\label{def:qf}
Q_f &=&\langle X_{i}^{\ast}X_{j}\rangle ,
\end{eqnarray}
for the nearest neighbour hopping and
\begin{eqnarray}
\label{def:qxprime}
Q_X' &=&\langle\sum_{\sigma}t_1\hat{f}_{ i\sigma}^\dag \hat{f}_{ j\sigma}\rangle , \\
\label{def:qfprime}
Q_f' &=&\langle X_{i}^{\ast}X_{j}\rangle
\end{eqnarray}
for the next nearest neighbour hopping.
Note that we consider only the half filled case which allowed us to set $\mu=h=0$.

For the spinon Hamiltonian,
\begin{eqnarray}\label{spinonHam}
&& H_f+H_f' \nonumber\\
&=&\sum_{\bold{k}}\Phi^{\dagger}_{\bold{k}}I\otimes(Q_f d_x\sigma_x+Q_f d_y\sigma_y+Q_f' d_z\sigma_z)\Phi_{\bold{k}}\nonumber\\
&=&\sum_{\bold{k}\sigma}(-\Sigma_{\bold{k}} \,{f_{\bold{k}\sigma}^{l\dagger}} {f_{\bold{k}\sigma}^l}
 + \Sigma_{\bold{k}} \, {f_{\bold{k}\sigma}^{u\dagger}} {f_{\bold{k}\sigma}^u}),
\end{eqnarray}
here $\Phi^{\dagger}_{\bold{k}}=( f_{\bold{k}\uparrow}^{a\dagger}, f_{\bold{k}\uparrow}^{b\dagger},
f_{\bold{k}\downarrow}^{a\dagger}, f_{\bold{k}\downarrow}^{b\dagger} )$. Thus, we obtain
the renormalized  spinon band structure
\begin{equation} \label{renorm-km-spec}
\Sigma_{\bold{k}} = \sqrt{ Q_f^{2}(d_x^2+d_y^2) +(Q_f'd_z)^2},
\end{equation}
which is very similar to that of spin-$\frac{1}{2}$ two-copy version of QAH model.

For the rotor Hamiltonian,
\begin{eqnarray}
H_X
&=&Q_X\sum_\bold{k} [-|g_1|{X_{\bold{k}}^{l\ast}}{X_{\bold{k}}^l} + |g_1|{X_{\bold{k}}^{u\ast}}{X_{\bold{k}}^u}\ ],\\
 H_X'
 &=& -Q_X' \sum_{\bold{k}} g_2(\bold{k})( {X_{\bold{k}}^{l\ast}} {X_{\bold{k}}^l} +
{X_{\bold{k}}^{u\ast}} {X_{\bold{k}}^u} )\ ,
\end{eqnarray}
the
Green function for the $X$ fields is written as
\begin{equation}\label{green:X}
G_X =  \frac{1}{\frac{\nu_n^2}{2U} + \rho + \xi_{\bold{k}}},
\end{equation}
here we ignore the upper band of rotor and consider only the lower band
\begin{equation}\label{def:xi_k}
\xi_{\bold{k}} = -Q_X |g_1(\bold{k})| - Q_X' g_2(\bold{k}) ,
\end{equation}
with $g_1(\bold{k})=\sum_{i=1}^{4}e^{-i\bold{k}\cdot \bs{\delta}_{i}}$,
$g_2(\bold{k})=\sum_{i=1}^{4}e^{-i\bold{k}\cdot \bs{\delta}_{i}'}$, and bosonic Matsubara frequency $\nu_{n}$.

To treat the constraint (\ref{constraint2}) on average,
we find a self-consistent equation:
\begin{eqnarray}\label{SCE1}
 1&=&\frac{1}{N}\sum_{\bold{k}}\frac{1}{\beta}\sum_{n}G_X(\bold{k},\nu_{n})\nonumber\\
 &=&\frac{\sqrt{2}U}{N}
 \sum_{\bold{k}}\frac{1}{\sqrt{\Delta_{X}^{2}+4U[\xi_{\bold{k}}-\min(\xi_{\bold{k}})]}}
\end{eqnarray}
by introducing the insulating gap of rotor
\begin{equation}
\label{introtorgap}
\Delta_X = 2\sqrt{U[\rho + \min{(\xi_{\bold{k}})}]}\ .
\end{equation}
If the phase transition from the Mott insulator to the superfluid of the rotor takes place, the rotor gap $\Delta_X$ must close. It indicates that
\begin{equation}\label{Uc-SOI}
U_{c_{1}}(t_1) = \frac{1}{2}\left[ \frac{1}{2N}\sum_{\bold{k}'} \frac{1}{\sqrt{ \xi_{\bold{k}} - \min(\xi_{\bold{k}}) }} \right]^{-2}\ ,
\end{equation}
which defines the critical interaction strength of Mott insulator, i.e when $U\leq U_{c_{1}}, \Delta_X=0$; $U>U_{c_{1}},\Delta_X>0$.

\begin{figure}[ht]
\includegraphics[width=1\columnwidth]{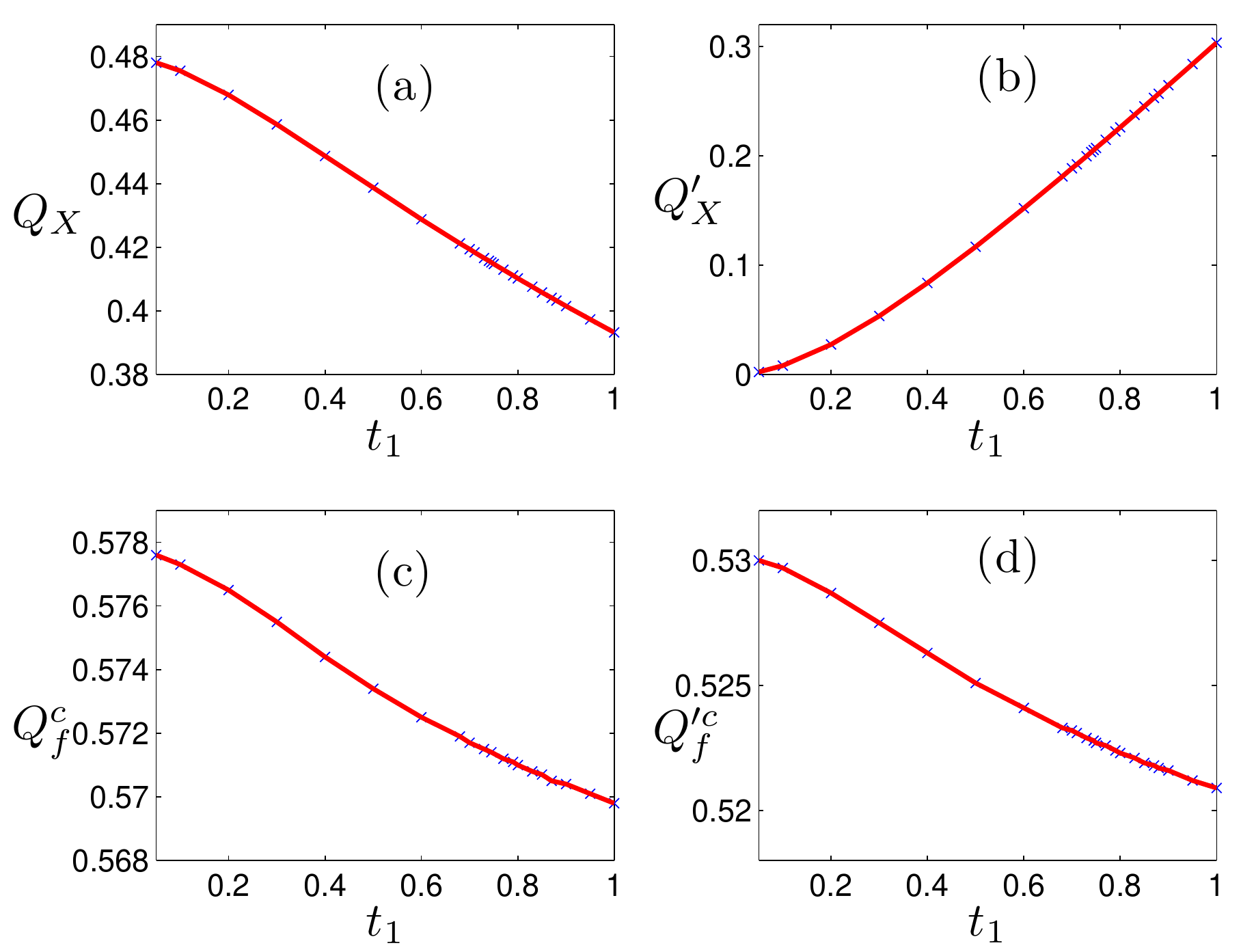}
\caption{(Color online)Numerical solutions of the mean-field parameters \eqref{def:qx}, \eqref{def:qf},
\eqref{def:qxprime} and \eqref{def:qfprime} ($t_0=1$) along the phase boundary $U_{c_{1}}$: (a) $Q_X(t_1)$,
(b)$Q'_X(t_1)$, (c)$Q_f^c(t_1)$, (d)$Q_f'^c(t_1)$.}
\label{qahcsltrans}
\end{figure}

After some detailed calculation (see appendix \ref{app1}), we may solve the mean-field equations corresponding to
\eqref{def:qx}, \eqref{def:qf}, \eqref{def:qxprime}, \eqref{def:qfprime} and \eqref{Uc-SOI} self-consistently along the phase boundary $U_{c_{1}}$. The mean-field parameters ($Q_X$, $Q_X'$, $Q_f^c$, $Q_f'^c$) as functions of next nearest neighbour hopping coefficient $t_1$ are plotted in Fig.\ref{qahcsltrans}. Here the hopping phase $\phi_0=\frac{\pi}{4}$ such that the flux $|\Phi|=\pi$ in each square plaquette in Fig. \ref{lattice}(b).

It is shown that along the phase transition from QAH state to CSL, with $t_1$ increasing from $0.05$ to $1$, $Q'_X$ increases from nearly zero to about $0.3$, but $Q_X$, $Q_f^c$ and $Q_f'^c$ decrease. Moreover, both $Q_f^c$ and $Q_f'^c$ decrease very slowly.

\section{Transition from CSL state to magnetically-ordered phase}\label{cslmag}

\begin{figure}[ht]
\includegraphics[width=1\columnwidth]{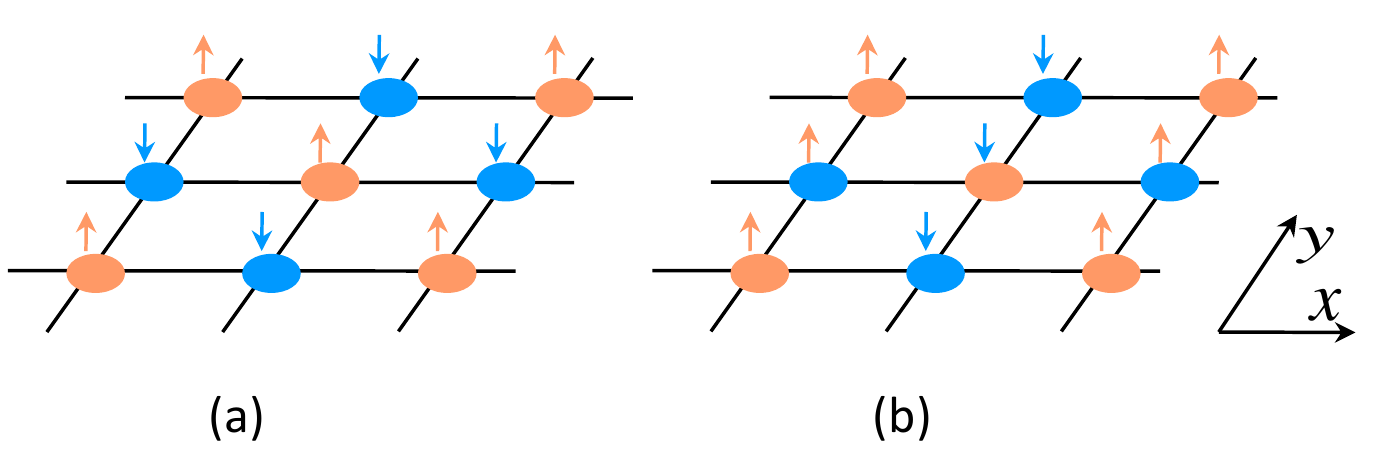}
\caption{(Color online) (a) The Neel order $m_1$, (b) The stripe order $m_2$. }
\label{neelstripe}
\end{figure}

Since the system may contain symmetry breaking orders at some parameter region, we may introduce magnetic order parameters $m_1$ and $m_2$
to describe Neel order and stripe order, respectively, which are two typical symmetry breaking orders. In the Neel order, the staggered spin order exists in the $x$ and $y$ directions [Fig.\ref{neelstripe}(a)]. While in the stripe order, the staggered spin order exists only in the $x$ or $y$ direction [Fig.\ref{neelstripe}(b)].

If we consider Neel state and stripe state at the same time, the square lattice can be divided into four sublattices
($A,B,C,D$), the area of unit cell is enlarged to be twice as large as the area of original unit cell without magnetic order.
Hence, the area of the corresponding FBZ is reduced to be half of area of the original FBZ.
The new FBZ is called reduced Brillouin zone (RBZ).
The sum over $\bold{k}$ on the RBZ is denoted as $\sum_{\bold{k}\in\,{\rm RBZ}}$, and the number of unit cells is $N_0$.
The magnetic order parameters are generally written as
\begin{equation}\label{nsorder1}
\langle n_{i\uparrow}-n_{i\downarrow} \rangle=\frac{1}{2}[(-1)^{i_{x}+i_{y}}m_1+(-1)^{i_x}m_2],
\end{equation}
with $i_{x}=\frac{x_i}{a}$, $i_y=\frac{y_i}{a}$.

\begin{figure}[ht]
\includegraphics[width=0.6\columnwidth]{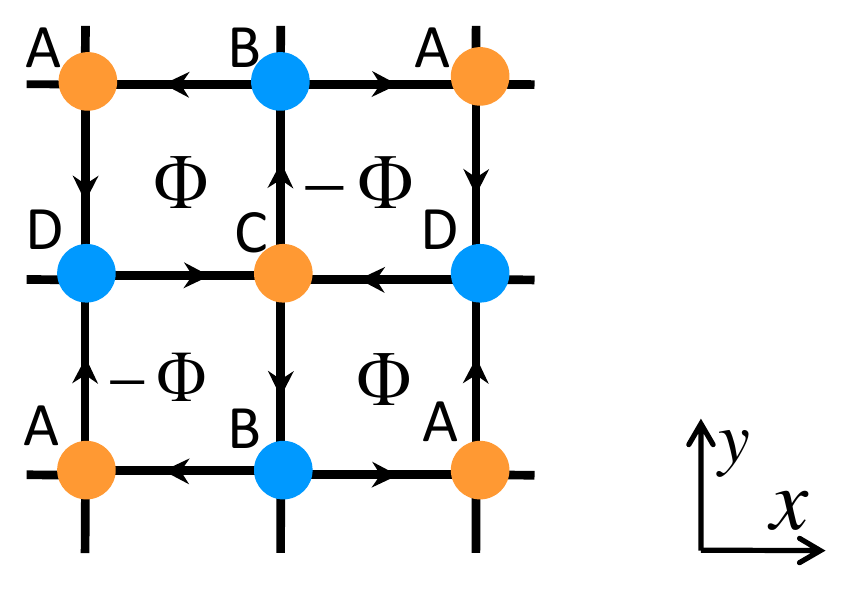}
\caption{(Color online) The square lattice containing four sublattices ($A,B,C,D$) for spinon because of magnetic orders. }
\label{latticemagorder}
\end{figure}

As mentioned in section \ref{srmf}, in the slave-rotor representation, the original fermion operator can be rewritten by using spinon operator $\hat{f}_{i\sigma}$ and rotor field $e^{i\theta_i}$: $\hat{c}_{i\sigma}=e^{i\theta_i}\hat{f}_{i\sigma}$.
The mean-field decomposed action takes the following form
\begin{eqnarray}
S_0&=&\int_0^\beta d\tau \Big[ \sum_{i\sigma} \hat{f}_{i\sigma}^\dagger \left( \partial_\tau - \mu + h_i \right)\hat{f}_{i\sigma}  + \sum_i \rho_i |X_i|^2\nonumber\\
& &+ \frac{1}{2U}\sum_i[(i\partial_\tau +i h_i)X_i^\ast][(-i\partial_\tau + ih_i)X_i]\nonumber\\
& &
+\sum_i \left( -h_i + \frac{h_i^2}{2U}\right)  +H_X +H_X'+H_f+H_f'\nonumber\\
& &+H_f''\Big]
\label{decomaction2}
\end{eqnarray}
with
\begin{eqnarray}
H_X&=&-\langle\sum_{\sigma}t_{ i j}e^{i\phi_{ i  j}}\hat{f}_{i\sigma}^{\dag}\hat{f}_{j\sigma}\rangle\sum_{\langle i j\rangle}X_{i}^{\ast}X_{j}+ {\rm c.c.}\nonumber\\
&=&-Q_{X_1}\sum_{i}X_{i}^{\ast}X_{i\pm1_x}\nonumber\\
&&-Q_{X_2}\sum_{i}X_{i}^{\ast}X_{i\pm1_y}+ {\rm c.c.} ,\label{magHx}\\
H_f&=&-Q_f\sum_{\langle i j\rangle\sigma}t_{ i j}e^{i\phi_{ i  j}} \hat{f}_{i\sigma}^{\dag} \hat{f}_{j\sigma}
+ {\rm H.c.} ,\label{magHf}\\
H_X'&=&-Q_X'\sum_{\langle\langle i j\rangle\rangle}X_{i}^{\ast}X_{j},\label{magHxprm}\\
H_f'&=&-Q_f'\sum_{\langle\langle i j\rangle\rangle}\sum_{\sigma}t_1\hat{f}_{ i\sigma}^\dag \hat{f}_{ j\sigma},\label{magHfprm}\\
H_f''&=&-\frac{U}{4} \sum_i[(-1)^{i_{x}+i_{y}}m_1+(-1)^{i_x}m_2]\nonumber\\& &
\cdot(\hat{f}_{i\uparrow}^\dagger \hat{f}_{i\uparrow}^{\phantom{\dagger}}-\hat{f}_{i\downarrow}^\dagger \hat{f}_{i\downarrow}^{\phantom{\dagger}})+\frac{UN_0}{4}(m_1^2+m_2^2)+c,\label{magHfpprm}
\end{eqnarray}
here the
next nearest neighbour hopping coefficients $t'_{a}=t'_{b}=t'_{c}=t'_{d}=t_{1}$
for $A,B,C,D$ sublattices,
$\frac{U}{4}\sum_{i}n_i^{2}=c$ is a constant, $n_i=n_{i\uparrow}+n_{i\downarrow}$.
We still fulfill the constraints \eqref{constraint1} and \eqref{constraint2}  with the Lagrange multipliers $h_i$ and $\rho_i$, respectively. And $h_{i}\equiv h$ and $\rho_i\equiv\rho$ are not local site-dependent
since \eqref{constraint1} and \eqref{constraint2} are treated on average \cite{florens-04prb035114,zhaopara,StephanK-10prb}.
The magnetic orders in Fig.\ref{neelstripe} break spin rotation symmetry and only affect spinon degrees of freedom, so the square lattice can be divided into four sublattices
($A,B,C,D$) for spinon (see Fig.\ref{latticemagorder})
, the sum over $\bold{k}$ is denoted as $\sum_{\bold{k}\in\,{\rm RBZ}}$. But for the rotor, the square lattice still has two sublattices ($A,B$), the sum over $\bold{k}$ is still on the original FBZ and denoted as $\sum_{\bold{k}}$.

The mean-field parameters associated with the decomposition are given by
\begin{eqnarray}
\label{def:magqx1}
Q_{X_1} &=&\langle\sum_{\sigma}t_{ i j}e^{i\phi_{ i  j}}\hat{f}_{i\sigma}^{b\dag}\hat{f}_{j\sigma}^{a}\rangle
= \langle\sum_{\sigma}t_{ i j}e^{i\phi_{ i  j}}\hat{f}_{i\sigma}^{d\dag}\hat{f}_{j\sigma}^{c}\rangle, \nonumber\\ \\
\label{def:magqx2}
Q_{X_2} &=&\langle\sum_{\sigma}t_{ i j}e^{i\phi_{ i  j}}\hat{f}_{i\sigma}^{d\dag}\hat{f}_{j\sigma}^{a}\rangle
= \langle\sum_{\sigma}t_{ i j}e^{i\phi_{ i  j}}\hat{f}_{i\sigma}^{b\dag}\hat{f}_{j\sigma}^{c}\rangle, \nonumber\\ \\
\label{def:magqf}
Q_f &=&\langle X_{i}^{\ast}X_{j}\rangle ,
\end{eqnarray}
for the nearest neighbour hopping and
\begin{eqnarray}
\label{def:magqxprime}
Q_X' &=&\langle\sum_{\sigma}t_1\hat{f}_{ i\sigma}^\dag \hat{f}_{ j\sigma}\rangle , \\
\label{def:magqfprime}
Q_f' &=&\langle X_{i}^{\ast}X_{j}\rangle
\end{eqnarray}
for the next nearest neighbour hopping.
We still have $\mu=h_i=0$  for half filled case.

The
spinon Hamiltonian in momentum space is written as
\begin{eqnarray}\label{magspinham}
&&H_f+H_f'+H_f''\nonumber\\
&=&\sum_{\bold{k}\in\,{\rm RBZ},\sigma}\Phi_{\bold{k}\sigma}^{\dagger}
\mathcal{H}_{\sigma}\Phi_{\bold{k}\sigma}+\frac{UN_0}{4}(m_1^2+m_2^2)+c\nonumber\\
&=&\sum_{\bold{k}\in\,{\rm RBZ},\sigma}(-\Sigma_{1}\hat{f}_{\bold{k}\sigma}^{l_{1}\dagger}\hat{f}_{\bold{k}\sigma}^{l_1}
-\Sigma_{2}\hat{f}_{\bold{k}\sigma}^{l_{2}\dagger}\hat{f}_{\bold{k}\sigma}^{l_2}\nonumber\\
&&
 +\Sigma_{2}\hat{f}_{\bold{k}\sigma}^{u_{1}\dagger}\hat{f}_{\bold{k}\sigma}^{u_1}
 +\Sigma_{1}\hat{f}_{\bold{k}\sigma}^{u_{2}\dagger}\hat{f}_{\bold{k}\sigma}^{u_2})\nonumber\\
&&+\frac{UN_0}{4}(m_1^2+m_2^2)+c,
\end{eqnarray}
here $\Phi_{\bold{k}\sigma}=( \hat{f}_{\bold{k}\sigma}^{a}, \hat{f}_{\bold{k}\sigma}^{b},
\hat{f}_{\bold{k}\sigma}^{c}, \hat{f}_{\bold{k}\sigma}^{d} )$, $\mathcal{H}_{\uparrow}$ and $\mathcal{H}_{\downarrow}$ are two $4\times4$ matrices
presented in appendix \ref{app1}.
We get
the renormalized mean-field free energy at $T=0$ for spinon sector
\begin{equation}\label{remffreeenergy}
\mathcal{F}=\sum_{\bold{k}\in\,{\rm RBZ}}(-2\Sigma_1-2\Sigma_2)+\frac{UN_0}{4}(m_1^2+m_2^2)+c.
\end{equation}

The self-consistent equations about magnetic orders $m_1$ and $m_2$ are equivalent to energy extreme conditions $\frac{\partial \mathcal{F}}{\partial m_1}=\frac{\partial \mathcal{F}}{\partial m_2}=0$.
By solving them we may determine critical interaction strength $U_{c_{2}}$ of magnetic order, i.e. when $U>U_{c_{2}}$, $m_1>0,m_2=0$ (Neel state) or $m_1=0, m_2>0$ (stripe state); when $U_{c_{1}}<U\leq U_{c_{2}}$, $m_1=m_2=0$ (CSL),
where $U_{c_1}$ is critical interaction of Mott insulator determined by Eq.\eqref{Uc-SOI}.

For the rotor Hamiltonian
\begin{eqnarray}
H_X &=&\sum_\bold{k} [-|g(\bold{k})|{X_{\bold{k}}^{l\ast}}{X_{\bold{k}}^l} + |g(\bold{k})|{X_{\bold{k}}^{u\ast}}{X_{\bold{k}}^u}\ ],
\label{rotorX}\\
H_X'&=& -Q_X' \sum_{\bold{k}} g_2(\bold{k})( {X_{\bold{k}}^{l\ast}} {X_{\bold{k}}^l} +
{X_{\bold{k}}^{u\ast}} {X_{\bold{k}}^u} )\ \label{rotorXprm},
\end{eqnarray}
where $g(\bold{k})=2[Q_{X_1}\cos (k_xa) +Q_{X_2}\cos (k_ya)]$,
$g_2(\bold{k})=\sum_{i=1}^{4}e^{-i\bold{k}\cdot \bs{\delta}_{i}'}$,
the
Green function for the $X$ fields is written as
\begin{equation}\label{green:magX}
G_X =  \frac{1}{\frac{\nu_n^2}{2U} + \rho + \xi_{\bold{k}}}
\end{equation}
where we consider the lower band of rotor
\begin{equation}\label{def:magxi_k}
\xi_{\bold{k}} = -2|Q_{X_1}\cos (k_xa) +Q_{X_2}\cos (k_ya)| - Q_X' g_2(\bold{k}) ,
\end{equation}
with bosonic Matsubara frequency $\nu_{n}$.

From average of the constraint (\ref{constraint2})
we have the following self-consistent equation along the phase boundary $U_{c_{2}}$
\begin{equation}\label{SCE1mag}
 1=\frac{1}{N}\sum_{\bold{k}}\frac{1}{\beta}\sum_{n}G_X(\bold{k},\nu_{n})=\frac{\sqrt{2U_{c_{2}}}}{2N}
 \sum_{\bold{k}}\frac{1}{\sqrt{\rho+\xi_{\bold{k}}}}.
\end{equation}
We can solve the equation to get Lagrange multiplier $\rho_c$.
It can be found $\rho_{c} > -\min{(\xi_{\bold{k}})}$, so the insulating gap of rotor $\Delta_X = 2\sqrt{U[\rho + \min{(\xi_{\bold{k}})}]}$ is nonzero.

\begin{figure}[ht]
\includegraphics[width=1\columnwidth]{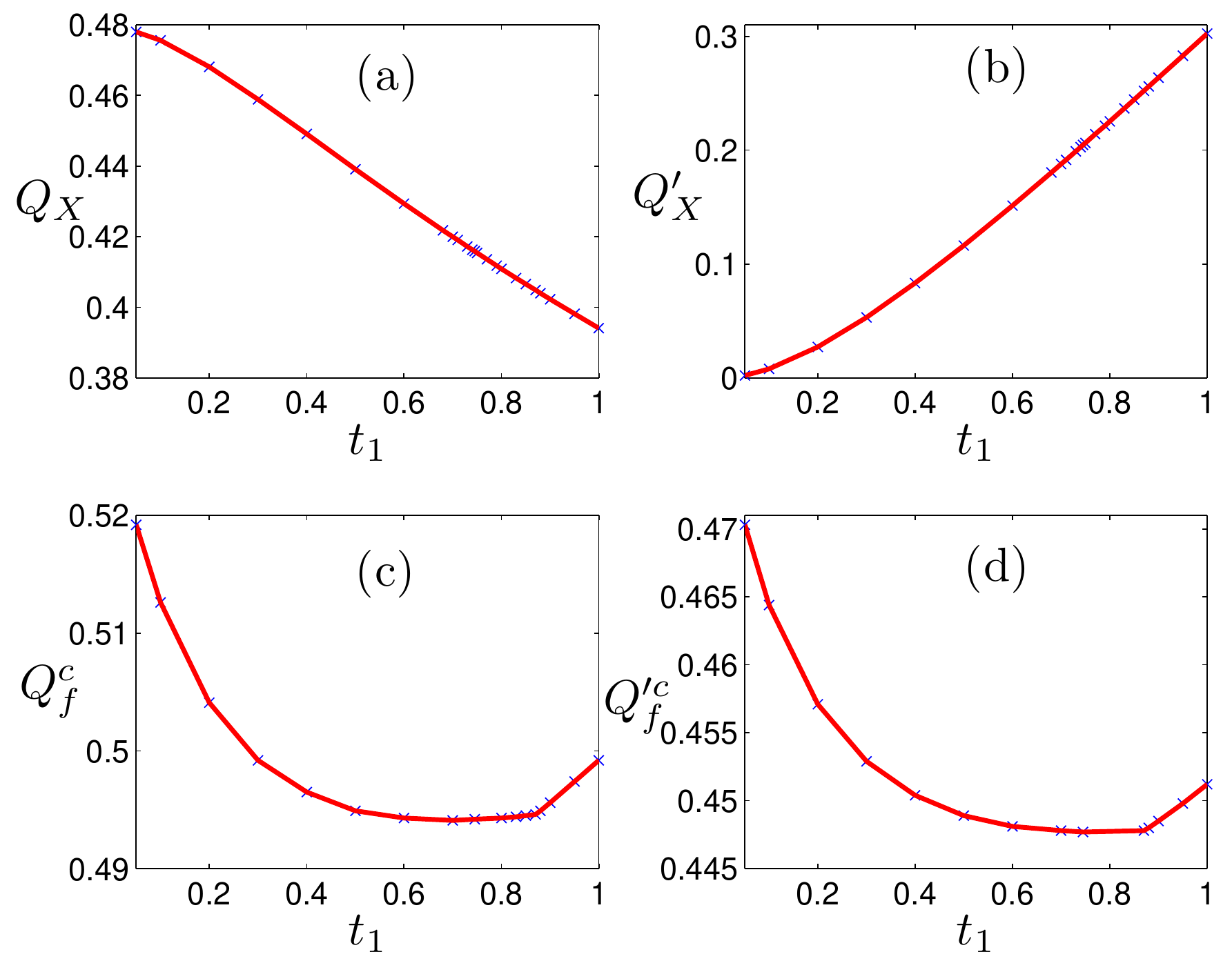}
\caption{(Color online) Numerical solutions of the mean-field parameters \eqref{def:magqx1}(\eqref{def:magqx2}), \eqref{def:magqf}, \eqref{def:magqxprime} and \eqref{def:magqfprime} ($t_0=1$) along the phase boundary $U_{c_{2}}$: (a) $Q_X(t_1)$,
(b)$Q'_X(t_1)$, (c)$Q_f^c(t_1)$, (d)$Q_f'^c(t_1)$.}
\label{csltomag}
\end{figure}

We still set hopping phase $\phi_0=\frac{\pi}{4}$. Equations $\frac{\partial \mathcal{F}}{\partial m_1}=\frac{\partial \mathcal{F}}{\partial m_2}=0$, \eqref{def:magqx1}, \eqref{def:magqx2}, \eqref{def:magqf},\eqref{def:magqxprime}, \eqref{def:magqfprime} and \eqref{SCE1mag}
can be solved self-consistently (see appendix \ref{app1}).
After careful numerical calculation we find \eqref{def:magqx1} is equal to \eqref{def:magqx2} along the phase transition from CSL to magnetically-ordered phase, so we have
$Q_{X_1}=Q_{X_2}\equiv Q_X$. We plot the mean-field parameters $Q_X$, $Q'_X$, $Q_f^c$ and $Q_f'^c$ as functions of next nearest neighbour hopping coefficient $t_1$ in Fig.\ref{csltomag}.

Comparing Fig.\ref{csltomag} with Fig.\ref{qahcsltrans}, with $t_1$ increasing from $0.05$ to $1$, along the phase transition from CSL to magnetically-ordered phase,  the behaviors of $Q_X$ and $Q'_X$ are very similar to those of $Q_X$ and $Q'_X$ along the phase transition from QAH state to CSL.
As for the behaviors of $Q_f^c$ and $Q_f'^c$ , different from  Figs.\ref{qahcsltrans} (c) and (d),  at the phase boundary of magnetic order, when $t_1<0.87$ both $Q_f^c$ and $Q_f'^c$ decrease at first, then vary very smoothly. But when $t_1>0.87$, they increase slowly. It indicates that magnetic order affects the spinon Hamiltonian, but not the rotor one.

\section{mean-field phase diagram of CSL in Fermi Hubbard model}\label{mfphase}

\begin{figure}[ht]
\includegraphics[width=1\columnwidth]{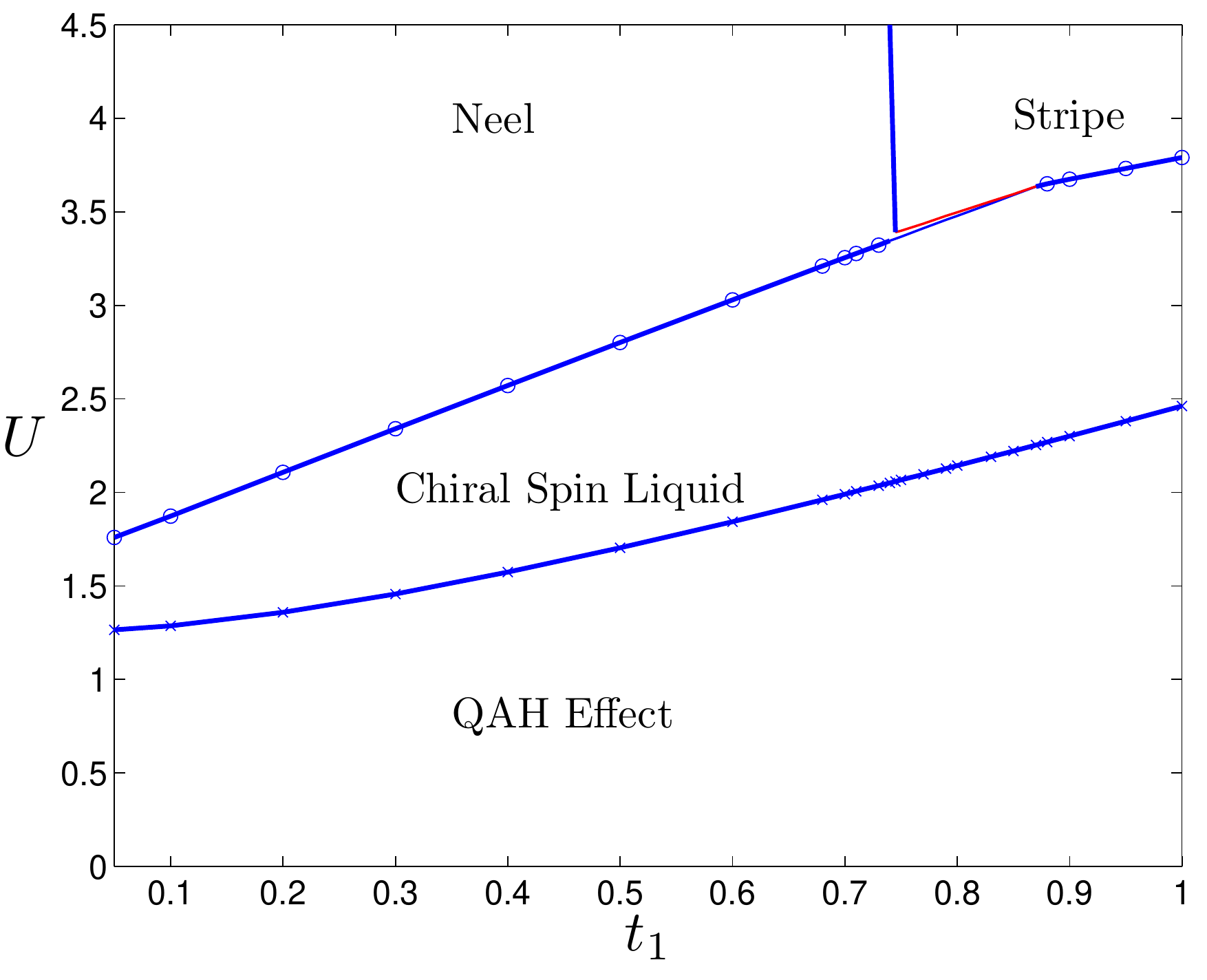}
\caption{(Color online) The slave-rotor mean-field phase diagram versus $t_1$ and Hubbard interaction $U$ of the optical Raman lattice. It has been set $t_0=1$.}
\label{slave-rotor}
\end{figure}

The self-consistent solutions of equations \eqref{Uc-SOI}, $\frac{\partial \mathcal{F}}{\partial m_1}=\frac{\partial \mathcal{F}}{\partial m_2}=0$ give critical interaction strength of CSL as functions of next nearest neighbour hopping coefficient $t_1$ in Fig.\ref{slave-rotor}.

It can be seen that when the strength of Hubbard interaction reaches a critical value $U_{c_1}$, the system will experience a phase transition from QAH state to CSL state. During this transition, the spin and charge are separated from each other. The charge degrees of freedom are frozen in the Mott insulator phase, while spinon degrees of freedom are described by the Hamiltonian \eqref{spinonHam}, which is exactly the spin-$\frac{1}{2}$ two-copy version of QAH model. This transition line corresponds to the lower boundary of CSL in Fig.\ref{slave-rotor}.

Since no magnetic order exists in the CSL phase, the CSL-magnetically-ordered phase transition will take place when interaction strength further increases and reaches another critical value $U_{c_2}$. This corresponds to the upper boundary of CSL in Fig.\ref{slave-rotor}. Along this transition line, when $0.05\leq t_1<0.87$, the CSL-Neel transition occurs first; when $0.87<t_1 \leq1$, the CSL-stripe transition occurs first. The three phases meet at a point ($t_1=0.87, U=3.635$).
When $0.745\leq t_1<0.87$, a Neel-stripe transition (the red line in Fig.\ref{slave-rotor}) will take place after CSL-Neel transition if interaction strength increases to a slightly larger critical value $U_{c_3}>U_{c_2}$, i.e. when $U_{c_2}<U\leq U_{c_3}$, $m_1\neq 0, m_2=0$ (Neel state); when $U>U_{c_3}$, $m_1=0, m_2\neq 0$ (stripe state).

\begin{figure}[ht]
\includegraphics[width=1\columnwidth]{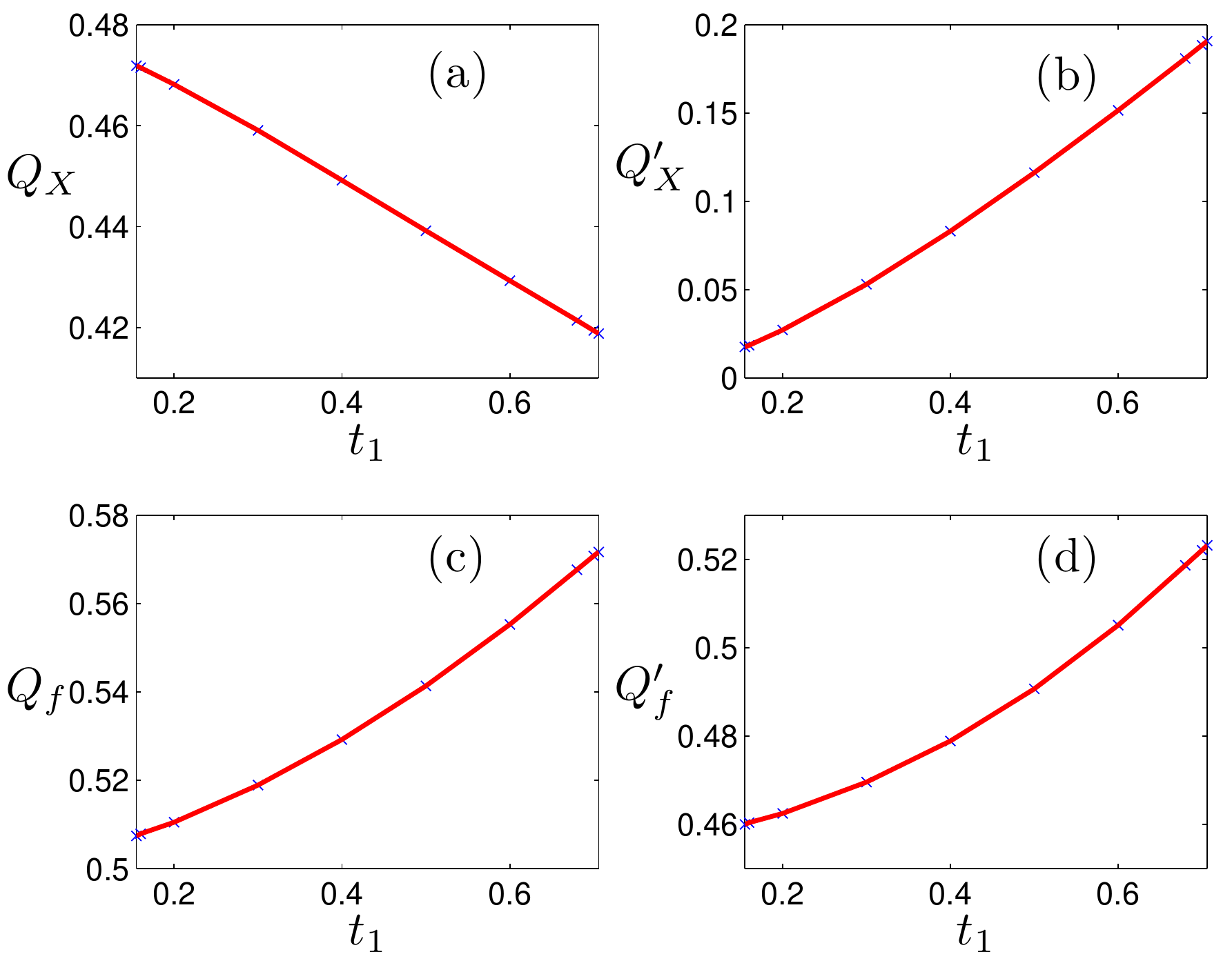}
\caption{(Color online)
For Hubbard interaction $U=2$, the evolution of mean-field parameters
in region $0.155\leq t_{1}\leq 0.706$, ($t_0=1$): (a) $Q_X(t_1)$,
(b)$Q'_X(t_1)$, (c)$Q_f(t_1)$, (d)$Q_f'(t_1)$.}
\label{qahcslmagu2}
\end{figure}

Furthermore, after solving Eq.\eqref{Uc-SOI} about critical interaction of Mott insulator, we find that for Hubbard interaction $U=2$, one phase transition from QAH
state to CSL state takes place at $t_1=0.706$, i.e. the insulating gap of rotor \eqref{introtorgap}
$\Delta_{X}=0$ when $U\leq 2$, but $\Delta_{X}>0$ when $U>2$.
And the solutions of $\frac{\partial \mathcal{F}}{\partial m_1}=\frac{\partial \mathcal{F}}{\partial m_2}=0$ with $\mathcal{F}$ given by \eqref{remffreeenergy} indicate that for $U=2$,
another phase transition
CSL-Neel takes place at $t_1=0.155$, i.e.
when $1.3222<U\leq 2$,
$m_1=m_2=0$ (CSL);
when
$U>2$, $m_1>0,m_2=0$ (Neel), where $U_{c_1}=1.3222$ is obtained by solving Eq.\eqref{Uc-SOI}.
When $0.155\leq t_{1}<0.706$ for $U=2$, the system is in CSL state.
These can be reflected in
the mean-field phase diagram Fig.\ref{slave-rotor} that for Hubbard interaction $U=2$, the system undergoes
Neel magnetic order phase, CSL phase and QAH phase.
We give the evolution of mean-field parameters
$Q_X$, $Q'_X$, $Q_f$ and $Q_f'$ in region $0.155\leq t_{1}\leq 0.706$ for $U=2$ by solving self-consistent equations \eqref{def:qx}, \eqref{def:qf}, \eqref{def:qxprime}, \eqref{def:qfprime} (see Eqs.
\eqref{SCE2}, \eqref{SCE2p},
\eqref{apSCE1csl}, \eqref{SCEcslqf} and \eqref{SCEcslqfprime} of appendix \ref{app1}).
In Fig.\ref{qahcslmagu2},
between phase transitions CSL-Neel and QAH-CSL,
as next nearest neighbour hopping coefficient $t_1$ increases, $Q'_X$, $Q_f$ and $Q_f'$ increase, but $Q_X$ decreases.


\section{chiral spin liquid phase in effective spin model}\label{spinmod}

It is well known that
when $U$ is large in the Hubbard interaction term \eqref{hubbard1}, each lattice site can not be doubly occupied, so the system is in a Mott insulator state.
We can derive an effective spin model by considering the Hilbert space with single occupied sites and treating the hopping terms as  perturbations.
In this section we will investigate the effective spin model and give more reasonable CSL phase diagram at mean-field level. Note that we can only give the phase boundary $U_c$ between CSL phase and magnetically-ordered phase by using the effective spin model. As for the transition from QAH state to CSL, we still use the results obtained from the slave-rotor formalism, i.e. the lower boundary $U_{c_{1}}$ of CSL in Fig.\ref{slave-rotor}.

\begin{figure}[ht]
\includegraphics[width=1\columnwidth]{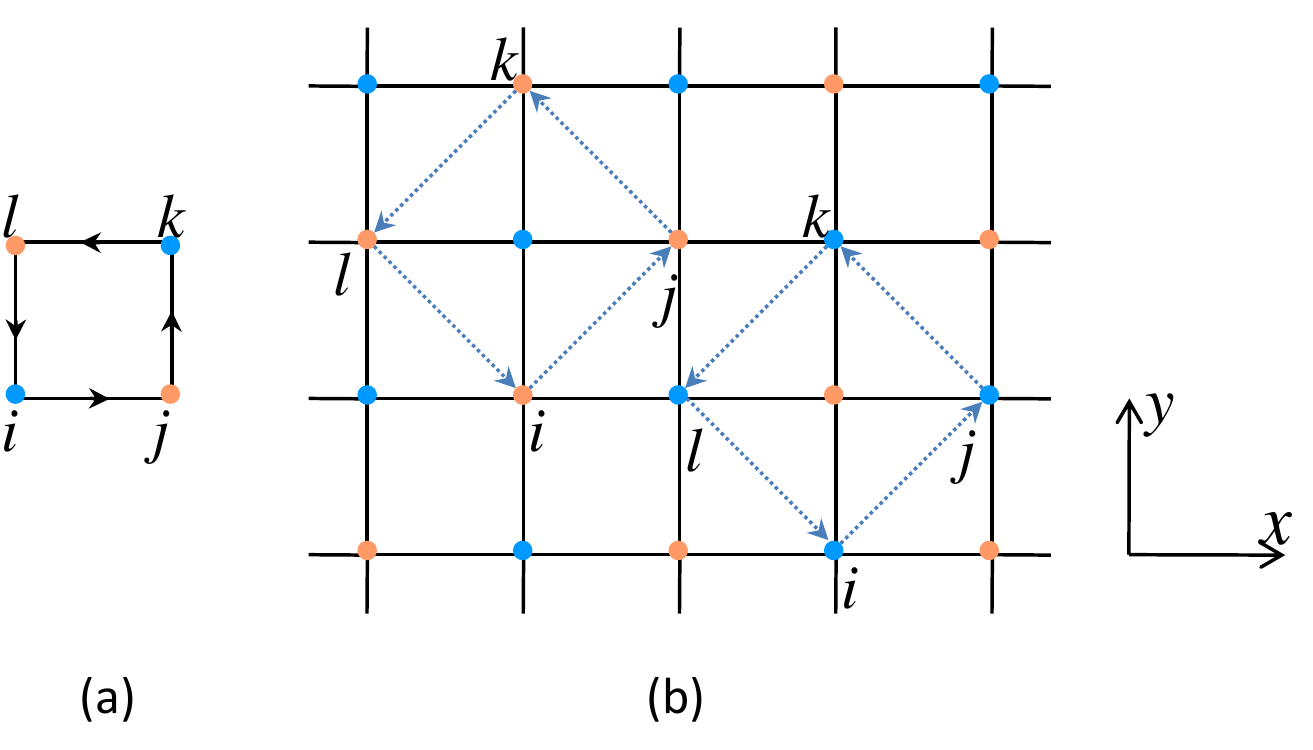}
\caption{(Color online)Schematic illustration of four spin interactions, each spin is interacted only once, (a)only nearest neighbour hopping, $\phi_{ijkl}=\pi$; (b)only next nearest neighbour hopping, $\phi_{ijkl}=0$.}
\label{fourspin1}
\end{figure}

As emphasized in section \ref{sec1}, the time-reversal symmetry of the system is broken in the CSL phase, so the effective spin model should at least include the terms up to third order perturbation expansions of $\frac{t_0}{U}$ and $\frac{t_1}{U}$, which correspond to interactions of three spins located at lattice sites of a closed minimum triangular loop. As for the fourth order expansions, they are exactly the four spin interaction terms. Relevant four spin interaction terms have also important effect on CSL phase, thus these terms should also be considered.
The derivation of effective spin model up to fourth order correction is presented in appendix \ref{app2}. Considering all spin configurations we can get the following effective Hamiltonian for the spin degrees of freedom with $t_0,t_1$ being small compared with $U$ \cite{MacDonald1988}
\begin{widetext}
\begin{eqnarray}\label{effspinham3rd}
H_{\rm eff}&=&\sum_{\langle ij \rangle}\frac{4t_0^2}{U}\mathbf{S}_{i}\cdot \mathbf{S}_{j}
 +\sum_{\langle\langle ij \rangle\rangle}\frac{4t_1^2}{U}\mathbf{S}_{i}\cdot \mathbf{S}_{j}+\sum_{ijk\in\bigtriangleup}\frac{24t_0^2 t_1}{U^2}\sin(\phi_{ijk})\mathbf{S}_{i}\cdot(\mathbf{S}_{j}\times\mathbf{S}_{k})\nonumber\\
 &&+\frac{t_{0}^{4}}{U^{3}}\sum_{ijkl\in \Box}\cos(\phi_{ijkl})\{80[(\mathbf{S}_{i}\cdot \mathbf{S}_{j})(\mathbf{S}_{k}\cdot\mathbf{S}_{l})+(\mathbf{S}_{i}\cdot \mathbf{S}_{l})(\mathbf{S}_{j}\cdot \mathbf{S}_{k})-(\mathbf{S}_{i}\cdot \mathbf{S}_{k})(\mathbf{S}_{j}\cdot \mathbf{S}_{l})] \nonumber\\
 && -4(\mathbf{S}_{i}\cdot \mathbf{S}_{j}+\mathbf{S}_{k}\cdot \mathbf{S}_{l}
      +\mathbf{S}_{j}\cdot \mathbf{S}_{k}+\mathbf{S}_{i}\cdot \mathbf{S}_{l}
      +\mathbf{S}_{i}\cdot \mathbf{S}_{k}+\mathbf{S}_{j}\cdot \mathbf{S}_{l})\}\nonumber\\
 &&+\frac{t_{1}^{4}}{U^{3}}\sum_{ijkl\in \diamond}\{80[(\mathbf{S}_{i}\cdot \mathbf{S}_{j})(\mathbf{S}_{k}\cdot\mathbf{S}_{l})+(\mathbf{S}_{i}\cdot \mathbf{S}_{l})(\mathbf{S}_{j}\cdot \mathbf{S}_{k})-(\mathbf{S}_{i}\cdot \mathbf{S}_{k})(\mathbf{S}_{j}\cdot \mathbf{S}_{l})] \nonumber\\
&& -4(\mathbf{S}_{i}\cdot \mathbf{S}_{j}+\mathbf{S}_{k}\cdot \mathbf{S}_{l}
      +\mathbf{S}_{j}\cdot \mathbf{S}_{k}+\mathbf{S}_{i}\cdot \mathbf{S}_{l}
      +\mathbf{S}_{i}\cdot \mathbf{S}_{k}+\mathbf{S}_{j}\cdot \mathbf{S}_{l})\}\nonumber\\
&& -\frac{16t_{0}^{4}}{U^{3}}\sum_{\langle ij\rangle}\mathbf{S}_{i}\cdot \mathbf{S}_{j}
-\frac{16t_{1}^{4}}{U^{3}}\sum_{\langle\langle ij\rangle\rangle}\mathbf{S}_{i}\cdot \mathbf{S}_{j}
+\frac{16t_{0}^{2}t_1^2}{U^{3}}\sum_{\langle ij\rangle}\mathbf{S}_{i}\cdot \mathbf{S}_{j}
+\frac{8t_{0}^{4}}{U^{3}}\sum_{\langle\langle ij \rangle\rangle}\mathbf{S}_{i}\cdot \mathbf{S}_{j},
\end{eqnarray}
\end{widetext}
where the spin operators at $i$th site are defined as:
$S^{x}_{i}=\frac{1}{2}(\hat{c}^{\dag}_{i\downarrow}\hat{c}_{i\uparrow}+\hat{c}^{\dag}_{i\uparrow}\hat{c}_{i\downarrow})$,
$S^{y}_{i}=\frac{i}{2}(\hat{c}^{\dag}_{i\downarrow}\hat{c}_{i\uparrow}-\hat{c}^{\dag}_{i\uparrow}\hat{c}_{i\downarrow})$,
$S^{z}_{i}=\frac{1}{2}(\hat{c}^{\dag}_{i\uparrow}\hat{c}_{i\uparrow}-\hat{c}^{\dag}_{i\downarrow}\hat{c}_{i\downarrow})$. This effective Hamiltonian is related to the parent Hamiltonians for CSL state constructed in terms of spin operators\cite{lslcft2012,parhamnabelcsl2014}.

As can be seen from \eqref{effspinham3rd}, the spin degree of freedom at each site is interacted with each other.
The first two terms reflect two spin interaction. The third term emerges only when the time-reversal symmetry of system is broken. The summation in the third term means that each set of $(i,j,k)$ consists of a minimum triangular.
$\phi_{ijk}$ is the Aharonov-Bohm phase acquired by hopping through the closed minimum triangular loop in anticlockwise direction $i\rightarrow j\rightarrow k\rightarrow i$. It can be verified that $\phi_{ijk}=\frac{\pi}{2}$ when $\phi_0=\frac{\pi}{4}$.

\begin{figure}[ht]
\includegraphics[width=1\columnwidth]{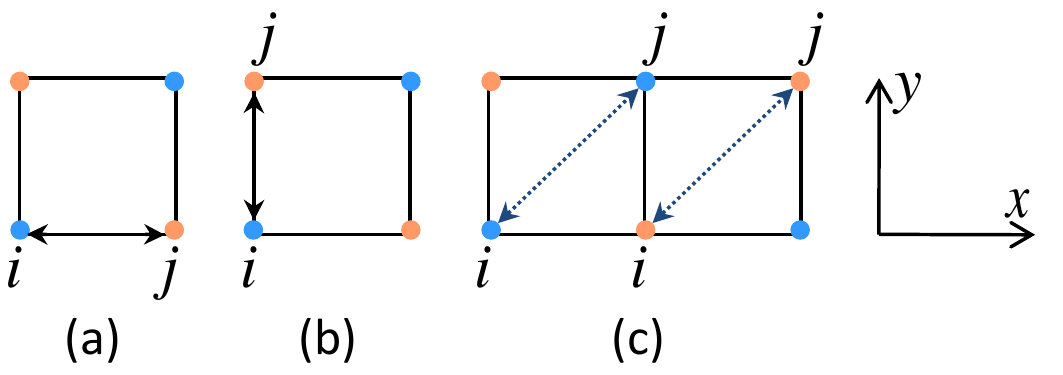}
\caption{(Color online) Schematic illustration of four spin interactions, each spin is interacted twice. (a) Nearest neighbour hopping along $x$ direction. (b) Nearest neighbour hopping along $y$ direction. (c) Next nearest neighbour hopping.}
\label{fourspin2}
\end{figure}

The fourth and fifth terms in \eqref{effspinham3rd} are about four spin interaction with four spins located at the four lattice sites $(i,j,k,l)$ of a plaquette, see Fig.\ref{fourspin1}. When spinons hop along the nearest neighbour bond in anticlockwise direction depicted in Fig.\ref{fourspin1}(a), a phase $\phi_0$ will be acquired, then it will give rise to a flux $\phi_{ijkl}=4\phi_0$ across each square plaquette. It is obvious that $\phi_{ijkl}=\pi$ if $\phi_0=\frac{\pi}{4}$. So the spinons experience a uniform U($1$) gauge field with the magnetic flux through each plaquette being $\pi$~\cite{Wen1989}.
In addition to hopping in anticlockwise direction, we should consider nearest neighbour hopping in clockwise direction to get the fourth term in \eqref{effspinham3rd}. However, the hopping events of spinons illustrated in Fig.\ref{fourspin1}(b) do not carry phase because the hopping is along next nearest neighbour bond. Such hopping events lead to the fifth term in \eqref{effspinham3rd}.

\begin{figure}[ht]
\includegraphics[width=1\columnwidth]{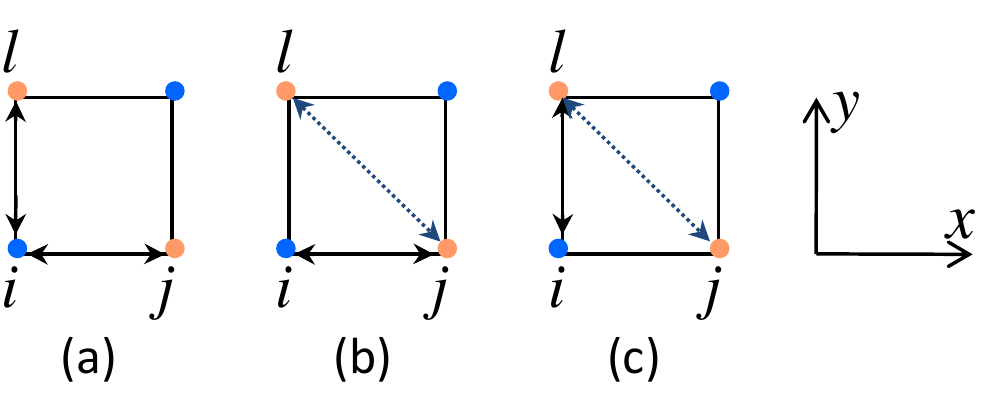}
\caption{(Color online) Schematic illustration of four spin interactions,
(a)spin at site $i$ is interacted twice; (b) spin at site $j$ is interacted twice; (c)spin at site $l$ is interacted twice.  }
\label{fourspin3}
\end{figure}

The four terms in the last line of \eqref{effspinham3rd} are also for four spin interactions. Different from those described above, they happen at two or three lattice sites. The first two terms correspond to four spin interactions at two neighbour sites, in which each site is interacted twice, see Fig.\ref{fourspin2}. While the last two terms correspond to four spin interactions at three sites of a minimum triangular. Among the three sites, one site is interacted twice, the other two sites are interacted only once, see Fig.\ref{fourspin3}.

We still use trial mean-field parameters introduced in Ref.\cite{xiongjun-16njp} to study different quantum phases.
The anyonic spinons $\hat{f}_i=(\hat{f}_{i\uparrow}, \hat{f}_{i\downarrow})^T$ and three Pauli matrices $\bs \sigma=(\sigma_x,\sigma_y,\sigma_z)$ can be used to represent the spin operator at $i$th site as $\bold S_i = \hat{f}_i^\dag{\bs \sigma\over2} \hat{f}_i$ under the particle number constraint $\hat{f}_i^\dag \hat{f}_i=1$.

At first, the system is assumed to have CSL phase,  so the mean-field parameters about hopping can be uesd to decouple the spin interactions:
\beq
\label{hoppara}
\chi_1
&=&+\langle \hat\chi_{ii+1_x}\rangle^*|_{i_x={\rm odd},i_y={\rm odd}}
\nonumber\\
&=&-\langle \hat\chi_{ii+1_x}\rangle|_{i_x={\rm even}
,i_y={\rm odd}}\nonumber\\
&=&+\langle \hat\chi_{ii+1_x}\rangle|_{i_x={\rm odd},i_y={\rm even}}
\nonumber\\
&=&-\langle \hat\chi_{ii+1_x}\rangle^*|_{i_x={\rm even}
,i_y={\rm even}}\nonumber\\
&=&+\langle \hat\chi_{ii+1_y}\rangle|_{i_x={\rm odd},i_y={\rm odd}}
\nonumber\\
&=&-\langle \hat\chi_{ii+1_y}\rangle^*|_{i_x={\rm odd},i_y={\rm even}}
\nonumber\\
&=&+\langle \hat\chi_{ii+1_y}\rangle^*|_{i_x={\rm even},i_y={\rm odd}}
\nonumber\\
&=&-\langle \hat\chi_{ii+1_y}\rangle|_{i_x={\rm even}
,i_y={\rm even}},
\nonumber\\
\chi_2
&=&
+\langle \hat\chi_{ii+1_x+1_y}\rangle=\chi_2^*
\eeq
with spinon hopping operator $\hat\chi_{ij} = \hat\chi_{ji}^\dag = \hat{f}_i^\dag \hat{f}_j$, $i_{x}=\frac{x_i}{a}$ and $i_y=\frac{y_i}{a}$. Here $\chi_2$ is assumed to be real and the hopping phase is only carried by $\chi_1$ for the CSL.

Furthermore, since the system may contain symmetry breaking orders at some parameter region, we need to use the following magnetic order parameters to describe the Neel order and stripe order, respectively
\begin{eqnarray}
M_n=(-1)^{i_x+i_y}\langle S^z_i\rangle, \ M_s = (-1)^{i_x}\langle S^z_i\rangle.
\end{eqnarray}
Now we can decouple the spin interactions by these trial mean-field parameters. The calculation is explicitly given in appendix \ref{app2}.

After decoupling the effective spin Hamiltonian \eqref{effspinham3rd} in terms of mean-field parameters $\chi_1$, $\chi_2$, $M_n$ and $M_s$, we obtain the matrix form of the Hamiltonian in real (coordinate) space. If the lattice system under investigation has $N=L_x \times L_y$ lattice sites, the Hamiltonian is a $2N \times 2N$ matrix. In calculation we use $N=16^{2}$. To minimize the free energy at $T=0$ with respect to the mean-field parameters $\chi_1$, $\chi_2$, $M_n$ and $M_s$,
we may determine the phase boundary
$U_c$ between the magnetic order and CSL state.


In the mean-field decoupling of the fourth term in \eqref{effspinham3rd}
\begin{eqnarray}\label{4spin1st}
&&-\{80[(\mathbf{S}_{i}\cdot \mathbf{S}_{j})(\mathbf{S}_{k}\cdot\mathbf{S}_{l})+(\mathbf{S}_{i}\cdot \mathbf{S}_{l})(\mathbf{S}_{j}\cdot \mathbf{S}_{k}) \nonumber\\
 &&-(\mathbf{S}_{i}\cdot \mathbf{S}_{k})(\mathbf{S}_{j}\cdot \mathbf{S}_{l})]
 -4(\mathbf{S}_{i}\cdot \mathbf{S}_{j}+\mathbf{S}_{k}\cdot \mathbf{S}_{l}
      +\mathbf{S}_{j}\cdot \mathbf{S}_{k} \nonumber\\
 && +\mathbf{S}_{i}\cdot \mathbf{S}_{l}
      +\mathbf{S}_{i}\cdot \mathbf{S}_{k}+\mathbf{S}_{j}\cdot \mathbf{S}_{l})\}\nonumber\\
 &=&5[\frac{4|\chi_1|^{4}e^{-i\pi}}{\langle\hat{\chi}_{li}\rangle}\hat{\chi}_{li}+
{\rm cyclic}(ijkl)-12|\chi_1|^{4}e^{-i\pi}+{\rm H.c.}]\nonumber\\
 &&-80[(-1)^{i_x+i_y}(M_{n}^{3}S^{z}_{i}-M_{n}^{3}S^{z}_{j}+M_{n}^{3}S^{z}_{k}-M_{n}^{3}S^{z}_{l})\nonumber\\
 &&-3M_n^4]
 -4(-1)^{i_x+i_y}(M_{n}S^{z}_{i}-M_{n}S^{z}_{j}+M_{n}S^{z}_{k}\nonumber\\
 &&-M_{n}S^{z}_{l})+8M_n^2,
\end{eqnarray}
here $j=i+1_x, k=i+1_x+1_y, l=i+1_y$, even if there is a $\pi$ flux in the hopping part, this term will reduce the
value of phase boundary $U_c$ obtained in Ref.\cite{xiongjun-16njp}) due to the $-80$ and $-4$ coefficients in the Neel order term. Therefore, in addition to \eqref{4spin1st}, the other four spin interaction terms in \eqref{effspinham3rd} are also very important for getting reasonable phase boundary.

\begin{figure}[ht]
\includegraphics[width=1\columnwidth]{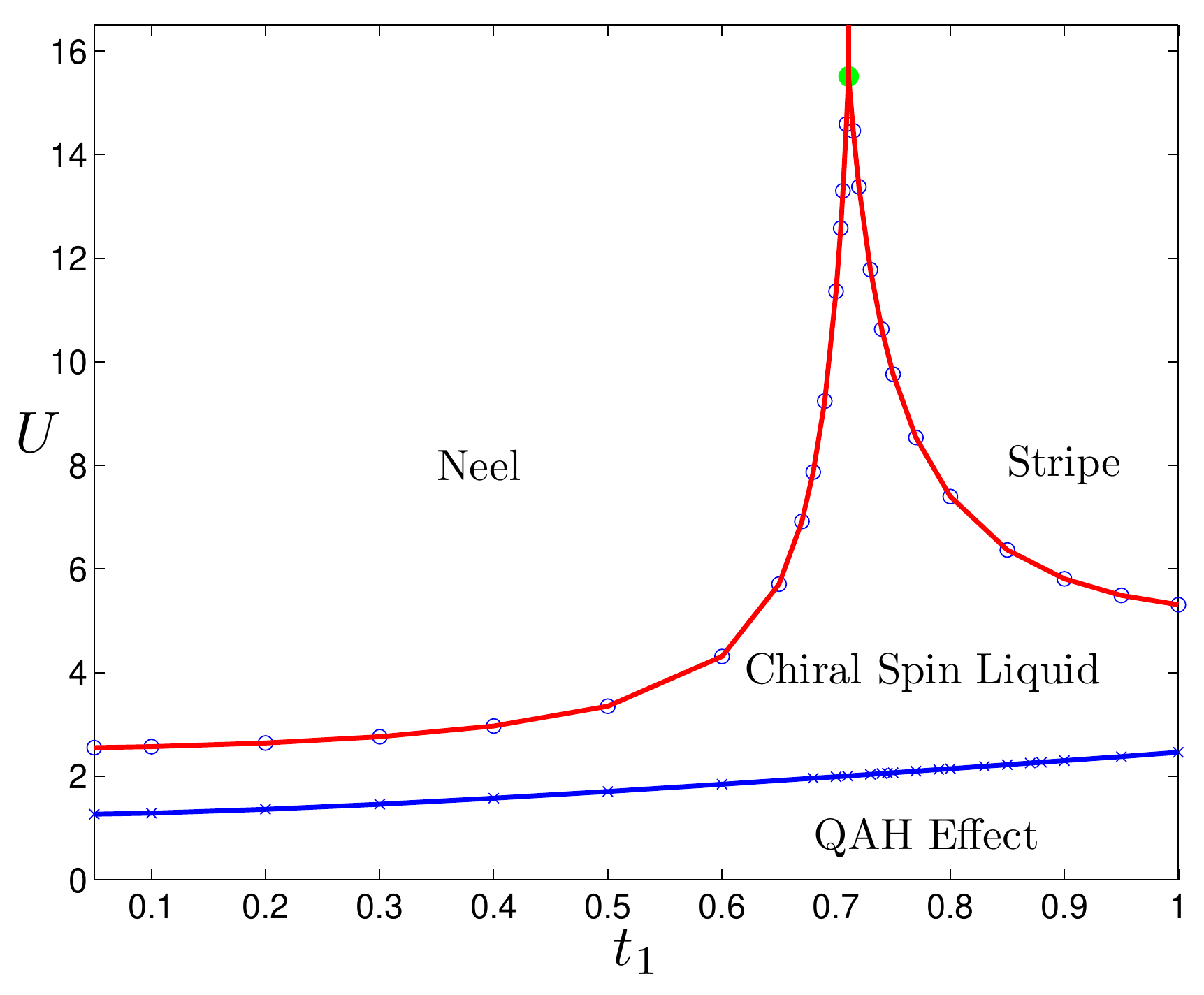}
\caption{(Color online) Spinon mean-field phase diagram ($t_0=1$) based on the effective spin model, in which the relevant four spin interactions are considered.}
\label{spinmodeldiag}
\end{figure}

In Fig.\ref{spinmodeldiag}, the CSL phase does appear at some parameter region:
when $U>U_{c}$, $\chi_1=0$, $\chi_2=0$, but $M_n\neq 0,M_s=0$ (Neel state) or $M_n=0, M_s\neq 0$ (stripe state);
when $U_{c_{1}}<U\leq U_{c}$, $\chi_1\neq 0$, $\chi_2\neq 0$, but $M_n=M_s=0$ (CSL).
With $t_1$ increasing in the region $0.05 \leq t_1 \leq 0.711$, the critical Hubbard interaction strength to reach the Neel order increases . However,
the critical Hubbard interaction strength to reach the stripe order decreases with $t_1$ increasing from $0.711$ to $1$. At $t_1=0.711$ the system becomes most frustrated (green point in Fig.\ref{spinmodeldiag}).

As emphasized before, the
spin chirality interaction
$\mathbf{S}_{i}\cdot(\mathbf{S}_{j}\times\mathbf{S}_{k})$
in effective Hamiltonian
\eqref{effspinham3rd} breaks time-reversal symmetry of system.
From expression \eqref{spinchiral} of $\mathbf{S}_{i}\cdot(\mathbf{S}_{j}\times\mathbf{S}_{k})$ in terms of spinon hopping operator $\hat\chi_{ij} = \hat\chi_{ji}^\dag = \hat{f}_i^\dag \hat{f}_j$
and the mean-field decomposition
\eqref{3spinchiral},
$\mathbf{S}_{i}\cdot(\mathbf{S}_{j}\times\mathbf{S}_{k})$
is proportional to
$\chi_{1}^2\chi_{2}-\chi_{1}^{\ast 2}\chi_{2}$, where $\chi_1$ and $\chi_2$ given by \eqref{hoppara}
are mean-field parameters about nearest neighbour hopping and next nearest neighbour hopping, respectively.
Since the nearest neighbour hopping phase $\phi_0=\frac{\pi}{4}$,
$\chi_1=|\chi_1|e^{i\frac{\pi}{4}}$.
In Table \ref{tab:chi1chi2}, we list $\chi_1$ and $\chi_2$ along the phase boundary $U_c$ for different next nearest neighbour hopping coefficient $t_1$.
This indicates that spin chirality interaction
$\mathbf{S}_{i}\cdot(\mathbf{S}_{j}\times\mathbf{S}_{k})$ is nonzero, thus CSL phase can be detected.

\begin{table}
\caption{For different $t_1$, mean-field parameters $\chi_1$ and $\chi_2$ along upper phase boundary of CSL in Fig.\ref{spinmodeldiag}.
}
\label{tab:chi1chi2}
\begin{tabular}{|c|c|c||c|c|c|}
\hline
$t_1$ & $\chi_1$ & $\chi_2$ &
$t_1$ & $\chi_1$ & $\chi_2$
\\
\hline

$0.05$
& $0.3917e^{i\frac{\pi}{4}}$
& $2.7796$

& $0.7$
& $0.4534e^{i\frac{\pi}{4}}$
& $0.1958$
\\
\hline

$0.1$
& $0.3933e^{i\frac{\pi}{4}}$
& $1.3887$

&$0.711$
& $0.4562e^{i\frac{\pi}{4}}$
& $0.1866$  \\
\hline

$0.2$
& $0.3986e^{i\frac{\pi}{4}}$
& $0.6913$

&$0.75$
& $0.4465e^{i\frac{\pi}{4}}$
&  $0.2154$ \\
\hline

$0.3$
& $0.4071e^{i\frac{\pi}{4}}$
& $0.4546$

&$0.8$
& $0.4384e^{i\frac{\pi}{4}}$
& $0.2345$  \\
\hline

$0.4$
& $0.42e^{i\frac{\pi}{4}}$
&  $0.3266$

&$0.85$
& $0.4321e^{i\frac{\pi}{4}}$
&  $0.2473$ \\
\hline

$0.5$
& $0.4416e^{i\frac{\pi}{4}}$
& $0.2306$

&$0.9$
& $0.4266e^{i\frac{\pi}{4}}$
& $0.2574$  \\
\hline

$0.6$
& $0.4503e^{i\frac{\pi}{4}}$
& $0.2052$

&$0.95$
& $0.4215e^{i\frac{\pi}{4}}$
& $0.2661$  \\
\hline

$0.65$
& $0.4493e^{i\frac{\pi}{4}}$
& $0.2078$

&$1$
& $0.4163e^{i\frac{\pi}{4}}$
& $0.274$  \\
\hline

\end{tabular}
\end{table}

Even if the four spin interaction terms in \eqref{effspinham3rd} do not break time-reversal symmetry of the system, they do have qualitative effect on the CSL phase diagram of effective Hamiltonian including only two and three spin interactions in Ref.\cite{xiongjun-16njp}. As in Fig.\ref{spinmodeldiag} we can see that the CSL phase is obtained in a broader region with $0.05\leq t_1 \leq 1$. The result is consistent with the phase diagram obtained in slave-rotor theory. We note that more accurate phase diagram can be obtained if all four spin interaction terms are taken into account, but it has only quantitative modification over the current phase diagram.

\section{Conclusion and Discussion}\label{sec:sum}
In this work, we have used two different mean-field approaches to investigate the CSL phase in an optical Raman lattice
with $\mathrm{U}(1)$ synthetic gauge flux. At first, we determine the phase boundary of CSL based on slave-rotor theory, which is applicable in strong Hubbard interacting regime.
When the Mott transition of charge takes place, the CSL phase appears. The spinon is separated from the charge, and the band structure of spinon is very similar to that of spin-$\frac{1}{2}$ two-copy version of QAH model with gapped bulk state and chiral gapless edge state. As a disordered phase without long-range spin order, the CSL preserves spin-rotational symmetry. Therefore no magnetic orders exist in the CSL, and the Mott insulator is a nonmagnetic insulator.

When Hubbard interaction strength $U$ is large, we manage to show CSL by spinon mean-field calculation based on an effective spin model derived from the Hubbard model and including not only two and three spin interaction terms, but also relevant four spin interaction terms. Our numerical results show that the CSL phase can be stabilized at strong magnetic frustrated system.

Our work
is an improved study on the previous work~\cite{xiongjun-16njp},
and
the critical Hubbard interaction strength of CSL depends on the mean-field approximation methods,
but
the phase diagrams obtained from the two different methods are consistent with each other, showing clear numerical evidence of the CSL phase obtained for cold atoms loaded to the improved optical Raman lattice setup~\cite{bz2018pra} which is of high experimental feasibility.
The mean-field approaches can be used to study other exotic topological phases for cold atoms of higher orbital bands and three-dimensional lattice systems.

\begin{acknowledgements}
We thank Zheng-Xin Liu, Ting Fung Jeffrey Poon, Sen Niu, Xin-Chi Zhou and  Bao-Zong Wang
for helpful discussions. This work was supported by National Key Research and Development Program of China (2021YFA1400900),  the Innovation Program for Quantum Science and Technology (Grant No. 2021ZD0302000), and the National Natural Science Foundation of China (Grants No. 11825401, No. 12261160368, and No. 11921005). 

\end{acknowledgements}


\appendix \section{Slave-rotor mean-field formalism}\label{app1}
In this appendix we show some detailed calculation about the slave-rotor mean-field approximation. At first we discuss the
transition from QAH to CSL.
For the spinon Hamiltonian \eqref{spinonHam}, we may get a
unitary transformation  between the sublattice bases $(\hat{f}_{\bold{k}\sigma}^{a},\hat{f}_{\bold{k}\sigma}^{b})$
and band bases $(\hat{f}_{\bold{k}\sigma}^{l},\hat{f}_{\bold{k}\sigma}^{u})$
\begin{equation}
\left(\begin{array}{c}\hat{f}_{\bold{k}\sigma}^{a}\\[5pt]
\hat{f}_{\bold{k}\sigma}^{b}\end{array}\right)=\left(\begin{array}{cc}\alpha_{-}&\alpha_{+}\\[5pt]
\beta_{-}&\beta_{+}\end{array}\right)
\left(\begin{array}{c}\hat{f}_{\bold{k}\sigma}^{l}\\[5pt]
\hat{f}_{\bold{k}\sigma}^{u}\end{array}\right). \label{unitrans1}
\end{equation}

By using the Fourier transformations:
\begin{eqnarray}
X_i(\tau)&=&\frac{1}{\sqrt{N}}\sum_{\bold k}e^{i \bold k \cdot \bold r_{i}}X_{\bold{k}}(\tau),\label{fouriertrans1}\\
X_{\bold{k}}(\tau)&=&\frac{1}{\sqrt{\beta}}\sum_{n}e^{-i\nu_{n}\tau}X_{\bold{k}}(\nu_{n}),\label{fouriertrans2}
\end{eqnarray}
we find the self-consistent equation from the constraint (\ref{constraint2})
\begin{eqnarray}\label{SCE1}
 1&=&\frac{1}{N}\sum_{\bold{k}}\frac{1}{\beta}\sum_{n}G_X(\bold{k},\nu_{n})\nonumber\\
 &=&\frac{\sqrt{2}U}{N}
 \sum_{\bold{k}}\frac{1}{\sqrt{\Delta_{X}^{2}+4U[\xi_{\bold{k}}-\min(\xi_{\bold{k}})]}}
\end{eqnarray}
In Eq.\,\eqref{SCE1} we perform the Matsubara sum on Green function (\ref{green:X}) at zero temperature (see Appendix D of \cite{StephanK-10prb}) and introduce the insulating gap of rotor
\begin{equation}
\label{introtorgapapp}
\Delta_X = 2\sqrt{U[\rho + \min{(\xi_{\bold{k}})}]}\ .
\end{equation}

The explicit form of insulating gap $\Delta_X$ should be determined by $\xi_{\bold k}$ and hence by $Q_X$ and $Q_X'$. The two mean-field parameters $Q_X$ and $Q_X'$ are the second and third self-consistent equations, respectively. They can be determined by using unitary transformation \eqref{unitrans1}.

We start with $Q_X$:
\begin{eqnarray}
&&\sum_{i=1}^{4}\langle\sum_{\sigma}t_{ i j}e^{i\phi_{ i  j}}\hat{f}_{i\sigma}^{b\dag}\hat{f}_{j\sigma}^{a}\rangle \nonumber\\
&=&
-\frac{1}{N}\sum_{\bold{k}\sigma}(d_x+id_y)\langle \hat{f}_{\bold{k}\sigma}^{b\dag}\hat{f}_{\bold{k}\sigma}^{a}\rangle
\nonumber\\
&=&-\frac{1}{N}\sum_{\bold{k}}2(d_x+id_y)\beta_{-}^{\ast}\alpha_{-}\langle \hat{f}^{l\dagger}_{\bold{k}\sigma}\hat{f}^{l}_{\bold{k}\sigma} \rangle\nonumber\\
&=&-\frac{1}{N}\sum_{\bold{k}}
2(d_x+id_y)\beta_{-}^{\ast}\alpha_{-},
\end{eqnarray}
where we assume $\langle \hat{f}^{l\dagger}_{\bold{k}\sigma}\hat{f}^{l}_{\bold{k}\sigma} \rangle=1$,
$\langle \hat{f}^{u\dagger}_{\bold{k}\sigma}\hat{f}^{u}_{\bold{k}\sigma} \rangle=0$ for $\sigma=\uparrow,\downarrow$  since the lower band of spinon is completely filled while the upper band is empty.
Due to the lattice symmetry, the sum over the four nearest neighbor sites, $\sum_{i=1}^4$, just
appears as a factor $4$ in the final expression. Thus we find the mean-field parameter
\beq
\label{SCE2}
Q_X
&=&
\langle\sum_{\sigma}t_{ i j}e^{i\phi_{ i  j}}\hat{f}_{i\sigma}^{b\dag}\hat{f}_{j\sigma}^{a}\rangle
\nonumber\\
&=&
-\frac{1}{4N} \sum_{\bold{k}}
2(d_x+id_y)\beta_{-}^{\ast}\alpha_{-}.
\eeq
Similarly,
\begin{eqnarray}
&&\sum_{i=1}^{4}\langle\sum_{\sigma}t_1\hat{f}_{ i\sigma}^\dag \hat{f}_{ j\sigma}\rangle
\nonumber\\
&=&-\frac{1}{N}\sum_{\bold{k}\sigma}d_z\langle \hat{f}_{\bold{k}\sigma}^{a\dag}\hat{f}_{\bold{k}\sigma}^{a}\rangle
=-\frac{1}{N}\sum_{\bold{k}\sigma}
(-d_z\langle \hat{f}_{\bold{k}\sigma}^{b\dag}\hat{f}_{\bold{k}\sigma}^{b}\rangle)
\nonumber\\
&=&-\frac{1}{N}\sum_{\bold{k}}2d_z|\alpha_{-}|^{2}
=-
\frac{1}{N}\sum_{\bold{k}}(-2d_z)
|\beta_{-}|^{2}.
\end{eqnarray}
Again the lattice symmetry is responsible for the fact that the sum over the four next nearest neighbor sites,
$\sum_{i=1}^4$, can be replaced by a factor $4$
\beq
\label{SCE2p}
Q_X'
&=&
\langle\sum_{\sigma}t_1\hat{f}_{ i\sigma}^{a\dag} \hat{f}_{ j\sigma}^{a}\rangle
=\langle\sum_{\sigma}t_{1}\hat{f}_{i\sigma}^{b\dag}
\hat{f}_{j\sigma}^{b}\rangle
\nonumber\\
&=&
-\frac{1}{4N}\sum_{\bold{k}}
2d_z|\alpha_{-}|^{2}
=-
\frac{1}{4N}\sum_{\bold{k}}
(-2d_z)|\beta_{-}|^{2}.
\nonumber\\
\eeq

The rotor spectrum $\xi_{\bold{k}}$ of Eq.\,\eqref{def:xi_k} is well defined and we can solve Eq.\,\eqref{SCE1}.
If the phase transition from the Mott insulator to the superfluid of the rotor takes place, the rotor gap $\Delta_X$ must close. It indicates that
\begin{equation}\label{apUc-SOI}
U_{c_{1}}(t_1) = \frac{1}{2}\left[ \frac{1}{2N}\sum_{\bold{k}'} \frac{1}{\sqrt{ \xi_{\bold{k}} - \min(\xi_{\bold{k}}) }} \right]^{-2}\ ,
\end{equation}
which defines the critical interaction strength of Mott insulator, i.e when $U\leq U_{c_{1}}, \Delta_X=0$; $U>U_{c_{1}},\Delta_X>0$.
The sum over $\bold{k}'$ means that formally the lowest bound corresponds to $\bold{k}\to\bold{k}_{\rm min}+\eta$, $\eta\ll 1$ \cite{StephanK-10prb}, and $\bold{k}_{\rm min}$ is wave vector associated with the minimum of $\xi_{\bold{k}}$. Hence, divergence in the sum can be avoided. This sum rule applies to \eqref{qfc} and \eqref{qfcprime}.

We have to study $Q_f$ and $Q_f'$ and their behaviors along the phase boundary $U_{c_{1}}(t_1)$.
By using Fourier transformations \eqref{fouriertrans1} and \eqref{fouriertrans2},
\begin{eqnarray}
\label{SCE3}
Q_f &=& \langle X_i^\ast X_j \rangle |_{{\rm nn.}} =
\frac{1}{N}\sum_{\bold{k}} e^{-i\bold{k}\cdot\bs{\delta}_\mu} \langle
X_{\bold{k}}^{b\ast} X_{\bold{k}}^a \rangle \nonumber\\
 &=& \frac{1}{N}\sum_{\bold{k}} \frac{|g_1|}{4}\frac{1}{\beta}\sum_n G_X(\bold{k},\nu_n)
\nonumber\\
&=&\frac{1}{N}\sum_{\bold{k}} \frac{|g_1|}{4}\frac{\sqrt{2}U}{2\sqrt{U\left(\rho + \xi_{\bold{k}}\right)}}\ ,
\end{eqnarray}
here $\bs{\delta}_\mu$ denotes one of the four nearest neighbor vectors in Fig.\ref{sqlattice},
$g_1(\bold{k})=\sum_{i=1}^{4}e^{-i\bold{k}\cdot \bs{\delta}_{i}}$. Along the transition line we have $\Delta_X=0$ and obtain
\begin{equation}\label{qfc}
Q_f^c(t_1) = \frac{\sqrt{2U_{c_{1}}(t_1)}}{8N}\sum_{\bold{k}'}
\frac{|g_1|}{\sqrt{\xi_{\bold{k}}- \min{(\xi_{\bold{k}})}}} \ .
\end{equation}

The last self-consistent equation determines $Q_f'$.
\begin{eqnarray}
\label{SCE3p}
 Q_f' &=& \langle X_i^\ast X_j \rangle |_{{\rm nnn.}}  \nonumber\\
&=& \frac{1}{N}\sum_{\bold{k}} e^{-i\bold{k}\cdot\bs{\delta}'_\mu} \langle X_{\bold{k}}^{(a/b)\ast} X_{\bold{k}}^{(a/b)}\rangle
\nonumber\\
  &=& \frac{1}{N}\sum_{\bold{k}} \frac{|g_2|}{4}\frac{1}{\beta}\sum_n G_X(\bold{k},\nu_n)
\nonumber\\
&=&\frac{1}{N}\sum_{\bold{k}} \frac{|g_2|}{4}\frac{\sqrt{2}U}{2\sqrt{U\left(\rho + \xi_{\bold{k}}\right)}}\ ,
\end{eqnarray}
with $\bs{\delta}'_\mu$ being one of the four next nearest neighbor vectors
in Fig.\ref{sqlattice},
$g_2(\bold{k})=\sum_{i=1}^{4}e^{-i\bold{k}\cdot \bs{\delta}_{i}'}$.
Thus we find $Q_f'$ along the Mott transition,
\beq\label{qfcprime}
Q_f'^c(t_1)= \frac{\sqrt{2U_{c_{1}}(t_1)}}{8N}\sum_{\bold{k}'}
\frac{|g_2|}{\sqrt{\xi_{\bold{k}}- \min{(\xi_{\bold{k}})}}}.
\eeq

If $U>U_{c_{1}},\Delta_X>0$, the system is in CSL state, the self-consistent equation from the constraint (\ref{constraint2})
is written as
\begin{equation}
\label{apSCE1csl}
 1=\frac{\sqrt{2U}}{2N}
 \sum_{\bold{k}}\frac{1}{\sqrt{\rho+\xi_{\bold{k}}}}.
\end{equation}
For $U>U_{c_{1}}$,
we may solve
the Lagrange multiplier $\rho$ ,
and the mean-field parameters $Q_f$ and $Q_f'$ can be calculated by $\rho$, $g_{1}(\bold{k})$ and $g_{2}(\bold{k})$ given above
\beq
\label{SCEcslqf}
Q_f = \frac{\sqrt{2U}}{8N}\sum_{\bold{k}}
\frac{|g_1|}{\sqrt{\xi_{\bold{k}}+\rho}} \ ,\\
\label{SCEcslqfprime}
Q_f' = \frac{\sqrt{2U}}{8N}\sum_{\bold{k}}
\frac{|g_2|}{\sqrt{\xi_{\bold{k}}+\rho}} \ .
\eeq

Secondly, we focus on the transition from CSL to magnetically-ordered phase. For the spinon Hamiltonian \eqref{magspinham},
the unitary transformation between the sublattice bases
$\Phi_{\bold{k}\sigma}=( \hat{f}_{\bold{k}\sigma}^{a}, \hat{f}_{\bold{k}\sigma}^{b},
\hat{f}_{\bold{k}\sigma}^{c}, \hat{f}_{\bold{k}\sigma}^{d} )$
and band bases $(\hat{f}^{l_1}_{\bold{k}\sigma},
\hat{f}^{l_2}_{\bold{k}\sigma},\hat{f}^{u_1}_{\bold{k}\sigma},\hat{f}^{u_2}_{\bold{k}\sigma})$
\beq
\left(\begin{array}{c}\hat f^{a}_{\bold{k}\sigma}\\[5pt]
\hat f^{b}_{\bold{k}\sigma}\\[5pt]
\hat f^{c}_{\bold{k}\sigma}\\[5pt]
\hat f^{d}_{\bold{k}\sigma}\end{array}\right)=\left(\begin{array}{cccc}\alpha^1_{\sigma-}&\alpha^2_{\sigma-}&\alpha^1_{\sigma+}&
\alpha^2_{\sigma+}\\[5pt]
\beta^1_{\sigma-}&\beta^2_{\sigma-}&\beta^1_{\sigma+}&
\beta^2_{\sigma+}\\[5pt]
\gamma^1_{\sigma-}&\gamma^2_{\sigma-}&\gamma^1_{\sigma+}&
\gamma^2_{\sigma+}\\[5pt]\delta^1_{\sigma-}&\delta^2_{\sigma-}&\delta^1_{\sigma+}&
\delta^2_{\sigma+}\end{array}\right)
\left(\begin{array}{c}\hat{f}^{l_1}_{\bold{k}\sigma}\\[5pt]
\hat{f}^{l_2}_{\bold{k}\sigma}\\[5pt]\hat{f}^{u_1}_{\bold{k}\sigma}\\[5pt]
\hat{f}^{u_2}_{\bold{k}\sigma}\end{array}\right)\label{unitrans2}
\eeq
can diagonalize
the two $4\times4$ matrices
$\mathcal{H}_{\uparrow}$
for $\sigma=\uparrow$
and $\mathcal{H}_{\downarrow}$
for $\sigma=\downarrow$,
\begin{widetext}
\beq
\mathcal{H}_{\uparrow}&=&\left(\begin{array}{cccc}
-\frac{U}{4}(m_1+m_2)  &2it_0 e^{i \frac{\pi}{4}}\sin(k_x a)Q_f&-4t_1 \cos (k_xa)\cos (k_ya)Q'_f&-2it_0 e^{-i \frac{\pi}{4}}\sin (k_y a)Q_f \\
-2i t_0e^{-i\frac{\pi}{4}}\sin( k_x a)Q_f&\frac{U}{4}(m_1+m_2) &2it_0 e^{i \frac{\pi}{4}}\sin( k_y a)Q_f&4t_1 \cos( k_xa)\cos (k_ya) Q'_f\\
-4t_1 \cos( k_xa)\cos( k_ya)Q'_f&-2it_0 e^{-i\frac{\pi}{4}}\sin(k_y a)Q_f&-\frac{U}{4}(m_1-m_2) &2it_0 e^{i\frac{\pi}{4}}\sin(k_x a)Q_f\\
2it_0 e^{i\frac{\pi}{4}}\sin( k_y a)Q_f &4t_1\cos(k_xa)\cos(k_ya)Q'_f&-2i t_0e^{-i\frac{\pi}{4}}\sin( k_x a) Q_f&\frac{U}{4}(m_1-m_2)
\end{array}\right)\ ,\nonumber\\  \label{spinonupham}
\eeq
\beq
\mathcal{H}_{\downarrow}&=&\left(\begin{array}{cccc}
\frac{U}{4}(m_1+m_2)  &2it_0 e^{i \frac{\pi}{4}}\sin(k_x a)Q_f&-4t_1 \cos (k_xa)\cos (k_ya)Q'_f&-2it_0 e^{-i \frac{\pi}{4}}\sin (k_y a)Q_f \\
-2i t_0e^{-i\frac{\pi}{4}}\sin( k_x a)Q_f&-\frac{U}{4}(m_1+m_2) &2it_0 e^{i \frac{\pi}{4}}\sin( k_y a)Q_f&4t_1 \cos( k_xa)\cos (k_ya) Q'_f\\
-4t_1 \cos( k_xa)\cos( k_ya)Q'_f&-2it_0 e^{-i\frac{\pi}{4}}\sin(k_y a)Q_f&\frac{U}{4}(m_1-m_2) &2it_0 e^{i\frac{\pi}{4}}\sin(k_x a)Q_f\\
2it_0 e^{i\frac{\pi}{4}}\sin( k_y a)Q_f &4t_1\cos(k_xa)\cos(k_ya)Q'_f&-2i t_0e^{-i\frac{\pi}{4}}\sin( k_x a) Q_f&-\frac{U}{4}(m_1-m_2)
\end{array}\right)\ ,\nonumber\\\label{spinondownham}
\eeq
\end{widetext}
here hopping phase $\phi_{0}=\frac{\pi}{4}$.
For the renormalized mean-field free energy at $T=0$ of spinon sector
obtained by above diagonalization,
\begin{equation}\label{remffreeenergyapp}
\mathcal{F}=\sum_{\bold{k}\in\,{\rm RBZ}}(-2\Sigma_1-2\Sigma_2)+\frac{UN_0}{4}(m_1^2+m_2^2)+c,
\end{equation}
the energy extreme conditions $\frac{\partial \mathcal{F}}{\partial m_1}=\frac{\partial \mathcal{F}}{\partial m_2}=0$
give the critical interaction strength $U_{c_{2}}$ of magnetic order.
The mean-field parameters $Q_{X_1}$ \eqref{def:magqx1}, $Q_{X_2}$ \eqref{def:magqx2} and
$Q_X'$ \eqref{def:magqxprime} along the phase boundary $U_{c_{2}}$ can be calculated by using the unitary transformation \eqref{unitrans2}.

We start with $Q_{X_{1}}$ and $Q_{X_{2}}$:
\beq
&&
\sum_{i=1}^{2}\langle\sum_{\sigma}t_{ i j}e^{i\phi_{ i  j}}\hat{f}_{i\sigma}^{b\dag}\hat{f}_{j\sigma}^{a}\rangle
\nonumber\\
&=&
-\frac{1}{N_{0}}\sum_{\mathbf{k}\in\mathrm{RBZ},\sigma}
[-2it_{0}e^{-i\frac{\pi}{4}}\sin(k_{x}a)]
\langle\hat{f}^{b\dag}_{\mathbf{k}\sigma}\hat{f}^{a}_{\mathbf{k}\sigma}\rangle
\nonumber\\
&=&
-\frac{1}{N_{0}}\sum_{\mathbf{k}\in\mathrm{RBZ},\sigma}
[-2it_{0}e^{-i\frac{\pi}{4}}\sin(k_{x}a)]
\nonumber\\
&&
\cdot
(\beta_{\sigma-}^{1\ast}\alpha_{\sigma-}^1+\beta_{\sigma-}^{2\ast}\alpha_{\sigma-}^2),
\eeq
\beq
&&
\sum_{i=1}^{2}\langle\sum_{\sigma}t_{ i j}e^{i\phi_{ i  j}}\hat{f}_{i\sigma}^{d\dag}\hat{f}_{j\sigma}^{c}\rangle
\nonumber\\
&=&
-\frac{1}{N_{0}}\sum_{\mathbf{k}\in\mathrm{RBZ},\sigma}
[-2it_{0}e^{-i\frac{\pi}{4}}\sin(k_{x}a)]
\langle\hat{f}^{d\dag}_{\mathbf{k}\sigma}\hat{f}^{c}_{\mathbf{k}\sigma}\rangle
\nonumber\\
&=&
-\frac{1}{N_{0}}\sum_{\mathbf{k}\in\mathrm{RBZ},\sigma}
[-2it_{0}e^{-i\frac{\pi}{4}}\sin(k_{x}a)]
\nonumber\\
&&
\cdot
(\delta_{\sigma-}^{1\ast}\gamma_{\sigma-}^1+\delta_{\sigma-}^{2\ast}\gamma_{\sigma-}^2),
\eeq
\beq
&&
\sum_{i=1}^{2}\langle\sum_{\sigma}t_{ i j}e^{i\phi_{ i  j}}\hat{f}_{i\sigma}^{d\dag}\hat{f}_{j\sigma}^{a}\rangle
\nonumber\\
&=&
-\frac{1}{N_{0}}\sum_{\mathbf{k}\in\mathrm{RBZ},\sigma}
2it_{0}e^{i\frac{\pi}{4}}\sin(k_{y}a)
\langle\hat{f}^{d\dag}_{\mathbf{k}\sigma}\hat{f}^{a}_{\mathbf{k}\sigma}\rangle
\nonumber\\
&=&
-\frac{1}{N_{0}}\sum_{\mathbf{k}\in\mathrm{RBZ},\sigma}
2it_{0}e^{i\frac{\pi}{4}}\sin(k_{y}a)
\nonumber\\
&&
\cdot
(\delta_{\sigma-}^{1\ast}\alpha_{\sigma-}^1+\delta_{\sigma-}^{2\ast}\alpha_{\sigma-}^2),
\eeq
\beq
&&
\sum_{i=1}^{2}\langle\sum_{\sigma}t_{ i j}e^{i\phi_{ i  j}}\hat{f}_{i\sigma}^{b\dag}\hat{f}_{j\sigma}^{c}\rangle
\nonumber\\
&=&
-\frac{1}{N_{0}}\sum_{\mathbf{k}\in\mathrm{RBZ},\sigma}
2it_{0}e^{i\frac{\pi}{4}}\sin(k_{y}a)
\langle\hat{f}^{b\dag}_{\mathbf{k}\sigma}\hat{f}^{c}_{\mathbf{k}\sigma}\rangle
\nonumber\\
&=&
-\frac{1}{N_{0}}\sum_{\mathbf{k}\in\mathrm{RBZ},\sigma}
2it_{0}e^{i\frac{\pi}{4}}\sin(k_{y}a)
\nonumber\\
&&
\cdot
(\beta_{\sigma-}^{1\ast}\gamma_{\sigma-}^1+\beta_{\sigma-}^{2\ast}\gamma_{\sigma-}^2).
\eeq
Thus the mean-field parameters
\beq
Q_{X_1} &=&\langle\sum_{\sigma}t_{ i j}e^{i\phi_{ i  j}}\hat{f}_{i\sigma}^{b\dag}\hat{f}_{j\sigma}^{a}\rangle
= \langle\sum_{\sigma}t_{ i j}e^{i\phi_{ i  j}}\hat{f}_{i\sigma}^{d\dag}\hat{f}_{j\sigma}^{c}\rangle
\nonumber\\
&=&
-\frac{1}{2N_{0}}\sum_{\mathbf{k}\in\mathrm{RBZ},\sigma}
[-2it_{0}
e^{-i\frac{\pi}{4}}\sin(k_{x}a)]
\nonumber\\
&&
\cdot(\beta_{\sigma-}^{1\ast}\alpha_{\sigma-}^1+\beta_{\sigma-}^{2\ast}\alpha_{\sigma-}^2)
\nonumber\\
&=&
-\frac{1}{2N_{0}}\sum_{\mathbf{k}\in\mathrm{RBZ},\sigma}
[-2it_{0}
e^{-i\frac{\pi}{4}}\sin(k_{x}a)]
\nonumber\\
&&
\cdot(\delta_{\sigma-}^{1\ast}\gamma_{\sigma-}^1+\delta_{\sigma-}^{2\ast}\gamma_{\sigma-}^2)
\label{SCE2mag1}
\eeq
and
\beq
Q_{X_2} &=&\langle\sum_{\sigma}t_{ i j}e^{i\phi_{ i  j}}\hat{f}_{i\sigma}^{d\dag}\hat{f}_{j\sigma}^{a}\rangle
= \langle\sum_{\sigma}t_{ i j}e^{i\phi_{ i  j}}\hat{f}_{i\sigma}^{b\dag}\hat{f}_{j\sigma}^{c}\rangle
\nonumber\\
&=&
-\frac{1}{2N_{0}}\sum_{\mathbf{k}\in\mathrm{RBZ},\sigma}
2it_{0}
e^{i\frac{\pi}{4}}\sin(k_{y}a)
\nonumber\\
&&
\cdot
(\delta_{\sigma-}^{1\ast}\alpha_{\sigma-}^1+\delta_{\sigma-}^{2\ast}\alpha_{\sigma-}^2)
\nonumber\\
\nonumber\\
&=&
-\frac{1}{2N_{0}}\sum_{\mathbf{k}\in\mathrm{RBZ},\sigma}
2it_{0}
e^{i\frac{\pi}{4}}\sin(k_{y}a)
\nonumber\\
&&
\cdot
(\beta_{\sigma-}^{1\ast}\gamma_{\sigma-}^1+\beta_{\sigma-}^{2\ast}\gamma_{\sigma-}^2),
\label{SCE2mag2}
\eeq
here for the lattice symmetry, the factor $2$ in the denominator denotes two nearest neighbour sites.

Similarly,
\beq
&&
\sum_{i=1}^{4}\langle\sum_{\sigma}t_{1}\hat{f}_{i\sigma}^{a\dag}\hat{f}_{j\sigma}^{c}\rangle
\nonumber\\
&=&
-\frac{1}{N_{0}}\sum_{\mathbf{k}\in\mathrm{RBZ},\sigma}
[-4t_{1}\cos(k_{x}a)\cos(k_{y}a)]
\langle\hat{f}^{a\dag}_{\mathbf{k}\sigma}\hat{f}^{c}_{\mathbf{k}\sigma}\rangle
\nonumber\\
&=&
-\frac{1}{N_{0}}\sum_{\mathbf{k}\in\mathrm{RBZ},\sigma}
[-4t_{1}\cos(k_{x}a)\cos(k_{y}a)]
\nonumber\\
&&
\cdot(\alpha_{\sigma-}^{1\ast}\gamma_{\sigma-}^1+\alpha_{\sigma-}^{2\ast}\gamma_{\sigma-}^2),
\eeq
\beq
&&
\sum_{i=1}^{4}\langle\sum_{\sigma}t_{1}\hat{f}_{i\sigma}^{b\dag}\hat{f}_{j\sigma}^{d}\rangle
\nonumber\\
&=&
-\frac{1}{N_{0}}\sum_{\mathbf{k}\in\mathrm{RBZ},\sigma}
4t_{1}\cos(k_{x}a)\cos(k_{y}a)
\langle\hat{f}^{b\dag}_{\mathbf{k}\sigma}\hat{f}^{d}_{\mathbf{k}\sigma}\rangle
\nonumber\\
&=&
-\frac{1}{N_{0}}\sum_{\mathbf{k}\in\mathrm{RBZ},\sigma}
4t_{1}\cos(k_{x}a)\cos(k_{y}a)
\nonumber\\
&&
\cdot(\beta_{\sigma-}^{1\ast}\delta_{\sigma-}^1+\beta_{\sigma-}^{2\ast}\delta_{\sigma-}^2),
\eeq
then the mean-field parameter
\beq
Q_X' &=&\langle\sum_{\sigma}t_1\hat{f}_{ i\sigma}^{a\dag} \hat{f}_{ j\sigma}^{c}\rangle
=\langle\sum_{\sigma}t_{1}\hat{f}_{i\sigma}^{b\dag}\hat{f}_{j\sigma}^{d}\rangle
\nonumber\\
&=&
-\frac{1}{4N_{0}}\sum_{\mathbf{k}\in\mathrm{RBZ},\sigma}
[-4t_{1}\cos(k_{x}a)\cos(k_{y}a)]
\nonumber\\
&&
\cdot(\alpha_{\sigma-}^{1\ast}\gamma_{\sigma-}^1+\alpha_{\sigma-}^{2\ast}\gamma_{\sigma-}^2)
\nonumber\\
&=&
-\frac{1}{4N_{0}}\sum_{\mathbf{k}\in\mathrm{RBZ},\sigma}
4t_{1}\cos(k_{x}a)\cos(k_{y}a)
\nonumber\\
&&
\cdot(\beta_{\sigma-}^{1\ast}\delta_{\sigma-}^1+\beta_{\sigma-}^{2\ast}\delta_{\sigma-}^2),
\label{SCE2pmag}
\eeq
here for the lattice symmetry, the factor $4$ in the denominator denotes four next nearest neighbour sites.
And $\langle \hat{f}^{l_{1}\dagger}_{\bold{k}\sigma}\hat{f}^{l_{1}}_{\bold{k}\sigma} \rangle=
\langle \hat{f}^{l_{2}\dagger}_{\bold{k}\sigma}\hat{f}^{l_{2}}_{\bold{k}\sigma}\rangle=1$,
$\langle \hat{f}^{u_{1}\dagger}_{\bold{k}\sigma}\hat{f}^{u_{1}}_{\bold{k}\sigma} \rangle=\langle \hat{f}^{u_{2}\dagger}_{\bold{k}\sigma}\hat{f}^{u_{2}}_{\bold{k}\sigma} \rangle=0$ for $\sigma=\uparrow,\downarrow$  since the lower band is completely filled while the upper band is empty.
The rotor spectrum $\xi_{\bold{k}}$ of Eq.\,\eqref{def:magxi_k} is well defined.

The  self-consistent equation along the phase boundary $U_{c_{2}}$
corresponding to the constraint (\ref{constraint2})
\begin{equation}\label{apSCE1mag}
 1=\frac{\sqrt{2U_{c_{2}}}}{2N}
 \sum_{\bold{k}}\frac{1}{\sqrt{\rho+\xi_{\bold{k}}}}.
\end{equation}
determines Lagrange multiplier $\rho_c$ along the phase boundary $U_{c_{2}}$.
Now we find the mean-field parameters $Q_f$ and $Q_f'$  along the transition line $U_{c_{2}}$,
\begin{eqnarray}\label{magqfc}
Q_f^c = \frac{\sqrt{2U_{c_{2}}}}{8N}\sum_{\bold{k}}
\frac{|g_1|}{\sqrt{\xi_{\bold{k}}+\rho_c}} \ ,\\
\label{magqfcprime}
Q_f'^c = \frac{\sqrt{2U_{c_{2}}}}{8N}\sum_{\bold{k}}
\frac{|g_2|}{\sqrt{\xi_{\bold{k}}+\rho_c}} \ ,
\end{eqnarray}
here $g_1(\bold{k})=\sum_{i=1}^{4}e^{-i\bold{k}\cdot \bs{\delta}_{i}}$
with
$\bs{\delta}_{i}$ being
one of four nearest neighbor vectors in Fig.\ref{sqlattice},
$g_2(\bold{k})=\sum_{i=1}^{4}e^{-i\bold{k}\cdot \bs{\delta}_{i}'}$
with
$\bs{\delta}_{i}'$ being
one of four next nearest neighbor vectors in Fig.\ref{sqlattice}.

\section{effective spin model}\label{app2}
In this appendix we will briefly present the derivation of effective spin Hamiltonian and how to decouple the spin interaction terms by
trial mean-field
parameters. We follow the procedure in \cite{MacDonald1988} and only give the corrections up to the fourth order. As for the fourth order terms, only the terms which have important physical meaning are considered. In other words, these fourth order terms can give qualitative modification on the mean-field phase diagram obtained from the effective spin Hamiltonian which has only expansions up to the third order \cite{xiongjun-16njp}.

Let's start from the total Hamiltonian
\begin{eqnarray}
H&=&H_0+H_{\rm int}\nonumber\\
&=&-\sum_{\langle i j\rangle\sigma}t_{ i j}e^{i\phi_{ i  j}} \hat{c}_{i\sigma}^\dag \hat{c}_{j\sigma}
-\sum_{\langle\langle i j\rangle\rangle}\sum_{\sigma}t_1\hat{c}_{i\sigma}^\dag \hat{c}_{ j\sigma}\nonumber\\
&&+U \sum_i n_{i\uparrow} n_{i\downarrow}.\label{tbHam4}
\end{eqnarray}
An effective spin model can be derived  by considering the perturbation expansions about $\frac{t_0}{U}$, $\frac{t_1}{U}$, with $t_0,t_1$ being small compared with $U$ at half-filled case.

By multiplying hopping term $H_0$ from the left by $1=n_{i\bar{\sigma}}+h_{i\bar{\sigma}}$
and from the right by $1=n_{j\bar{\sigma}}+h_{j\bar{\sigma}}$, we may rewrite Hamiltonian \eqref{tbHam4} in the following form
\begin{eqnarray}
H&=&T_{0}+H_{\rm int}+T_{+} +T_{-}, \nonumber\\
T_0&=&-\sum_{\langle ij\rangle\sigma}t_{ij}e^{i\phi_{ij}}(h_{i\bar{\sigma}}\hat{c}^\dagger_{i\sigma}\hat{c}_{j\sigma}h_{j\bar{\sigma}}
+n_{i\bar{\sigma}}\hat{c}^\dagger_{i\sigma}\hat{c}_{j\sigma}n_{j\bar{\sigma}}) \nonumber\\
&& -\sum_{\langle\langle ij \rangle\rangle}\sum_{\sigma}t_{1}(h_{i\bar{\sigma}}\hat{c}^\dagger_{i\sigma}\hat{c}_{j\sigma}h_{j\bar{\sigma}}
+n_{i\bar{\sigma}}\hat{c}^\dagger_{i\sigma}\hat{c}_{j\sigma}n_{j\bar{\sigma}}),\label{Tzero}\nonumber\\ \\
T_+&=&-\sum_{\langle ij\rangle\sigma}t_{ij}e^{i\phi_{ij}}n_{i\bar{\sigma}}\hat{c}^\dagger_{i\sigma}\hat{c}_{j\sigma}h_{j\bar{\sigma}}\nonumber\\
&&-
\sum_{\langle\langle ij \rangle\rangle}\sum_{\sigma}t_{1}n_{i\bar{\sigma}}\hat{c}^\dagger_{i\sigma}\hat{c}_{j\sigma}h_{j\bar{\sigma}},\label{Tplus}\\
T_-&=&-\sum_{\langle ij\rangle\sigma}t_{ij}e^{i\phi_{ij}}h_{i\bar{\sigma}}\hat{c}^\dagger_{i\sigma}\hat{c}_{j\sigma}n_{j\bar{\sigma}}\nonumber\\
&&-
\sum_{\langle\langle ij \rangle\rangle}\sum_{\sigma}t_{1}h_{i\bar{\sigma}}\hat{c}^\dagger_{i\sigma}\hat{c}_{j\sigma}n_{j\bar{\sigma}},\label{Tminus}
\end{eqnarray}
where the spin indices $\sigma=\uparrow,\downarrow$, $\bar{\sigma}$ is up for $\sigma$ down and down for $\sigma$ up, $n_{i\sigma}=\hat{c}^\dagger_{i\sigma}\hat{c}_{i\sigma}$ and $h_{i\sigma}=1-n_{i\sigma}$. $T_+$ describes the process of increasing one doubly-occupied site, $T_-$ describes the process of decreasing one doubly-occupied site and $T_0$ leaves the number of doubly-occupied sites unchanged.

The effective Hamiltonian can be obtained by a canonical transformation \cite{MacDonald1988}:
\begin{eqnarray}
H_{\rm eff}&=&e^{S}He^{-S} \nonumber\\
&=&H+[S,H]+\frac{1}{2}[S,[S,H]]+\frac{1}{3!}[S,[S,[S,H]]]\nonumber\\
&&+\frac{1}{4!}[S,[S,[S,[S,H]]]]+\cdots.
\end{eqnarray}
In the effective Hamiltonian the hopping events which increase or decrease doubly-occupied sites must be prohibited since the ground state of the system has no doubly-occupied sites. By using commutation relations $[T_{\pm},H_{\rm int}]=\mp UT_{\pm}$, $[T_0,H_{\rm int}]=0$, we can get the effective Hamiltonian up to the fourth order corrections\cite{MacDonald1988}:
\begin{eqnarray}
H_{\rm eff}&=&-\frac{1}{U}T_{-}T_{+}+\frac{1}{U^{2}}T_{-}T_{0}T_{+}
-\frac{1}{U^{3}}T_{-}T_{0}T_{0}T_{+}\nonumber\\
&&+\frac{1}{U^{3}}T_{-}T_{+}T_{-}T_{+}-\frac{1}{2U^{3}}T_{-}T_{-}T_{+}T_{+}.\label{effham}
\end{eqnarray}

If all spin configurations at half-filled case are taken into account, we can reach the following effective Hamiltonian for the spin degree of freedom
\begin{widetext}
\begin{eqnarray}\label{effspinham3rdapp}
H_{\rm eff}&=&\sum_{\langle ij \rangle}\frac{4t_0^2}{U}\mathbf{S}_{i}\cdot \mathbf{S}_{j}
 +\sum_{\langle\langle ij \rangle\rangle}\frac{4t_1^2}{U}\mathbf{S}_{i}\cdot \mathbf{S}_{j}+\sum_{ijk\in\bigtriangleup}\frac{24t_0^2 t_1}{U^2}\sin(\phi_{ijk})\mathbf{S}_{i}\cdot(\mathbf{S}_{j}\times\mathbf{S}_{k})\nonumber\\
 &&+\frac{t_{0}^{4}}{U^{3}}\sum_{ijkl\in \Box}\cos(\phi_{ijkl})\{80[(\mathbf{S}_{i}\cdot \mathbf{S}_{j})(\mathbf{S}_{k}\cdot\mathbf{S}_{l})+(\mathbf{S}_{i}\cdot \mathbf{S}_{l})(\mathbf{S}_{j}\cdot \mathbf{S}_{k})-(\mathbf{S}_{i}\cdot \mathbf{S}_{k})(\mathbf{S}_{j}\cdot \mathbf{S}_{l})] \nonumber\\
 && -4(\mathbf{S}_{i}\cdot \mathbf{S}_{j}+\mathbf{S}_{k}\cdot \mathbf{S}_{l}
      +\mathbf{S}_{j}\cdot \mathbf{S}_{k}+\mathbf{S}_{i}\cdot \mathbf{S}_{l}
      +\mathbf{S}_{i}\cdot \mathbf{S}_{k}+\mathbf{S}_{j}\cdot \mathbf{S}_{l})\}\nonumber\\
 &&+\frac{t_{1}^{4}}{U^{3}}\sum_{ijkl\in \diamond}\{80[(\mathbf{S}_{i}\cdot \mathbf{S}_{j})(\mathbf{S}_{k}\cdot\mathbf{S}_{l})+(\mathbf{S}_{i}\cdot \mathbf{S}_{l})(\mathbf{S}_{j}\cdot \mathbf{S}_{k})-(\mathbf{S}_{i}\cdot \mathbf{S}_{k})(\mathbf{S}_{j}\cdot \mathbf{S}_{l})] \nonumber\\
&& -4(\mathbf{S}_{i}\cdot \mathbf{S}_{j}+\mathbf{S}_{k}\cdot \mathbf{S}_{l}
      +\mathbf{S}_{j}\cdot \mathbf{S}_{k}+\mathbf{S}_{i}\cdot \mathbf{S}_{l}
      +\mathbf{S}_{i}\cdot \mathbf{S}_{k}+\mathbf{S}_{j}\cdot \mathbf{S}_{l})\}\nonumber\\
&& -\frac{16t_{0}^{4}}{U^{3}}\sum_{\langle ij\rangle}\mathbf{S}_{i}\cdot \mathbf{S}_{j}
-\frac{16t_{1}^{4}}{U^{3}}\sum_{\langle\langle ij\rangle\rangle}\mathbf{S}_{i}\cdot \mathbf{S}_{j}
+\frac{16t_{0}^{2}t_1^2}{U^{3}}\sum_{\langle ij\rangle}\mathbf{S}_{i}\cdot \mathbf{S}_{j}
+\frac{8t_{0}^{4}}{U^{3}}\sum_{\langle\langle ij \rangle\rangle}\mathbf{S}_{i}\cdot \mathbf{S}_{j},
\end{eqnarray}
\end{widetext}
where the spin operators at $i$th site are defined as:
$S^{x}_{i}=\frac{1}{2}(\hat{c}^{\dag}_{i\downarrow}\hat{c}_{i\uparrow}+\hat{c}^{\dag}_{i\uparrow}\hat{c}_{i\downarrow})$,
$S^{y}_{i}=\frac{i}{2}(\hat{c}^{\dag}_{i\downarrow}\hat{c}_{i\uparrow}-\hat{c}^{\dag}_{i\uparrow}\hat{c}_{i\downarrow})$,
$S^{z}_{i}=\frac{1}{2}(\hat{c}^{\dag}_{i\uparrow}\hat{c}_{i\uparrow}-\hat{c}^{\dag}_{i\downarrow}\hat{c}_{i\downarrow})$.

Obviously, the time-reversal symmetry of system is broken due to the appearance of
the third term. The summation in the third term means that each set of $(i,j,k)$ consists of a minimum triangular.
$\phi_{ijk}$ is the Aharonov-Bohm phase acquired by hopping through the closed minimum triangular loop in anticlockwise direction $i\rightarrow j\rightarrow k\rightarrow i$. It can be verified that $\phi_{ijk}=\frac{\pi}{2}$ when $\phi_0=\frac{\pi}{4}$.

To study different phases we should introduce mean-field parameters. At first we use the anyonic spinons $\hat{f}_i=(\hat{f}_{i\uparrow}, \hat{f}_{i\downarrow})^T$ to represent the spin operator as $\bold S_i = \hat{f}_i^\dag{\bs \sigma\over2} \hat{f}_i$ as long as the particle number $\hat{f}_i^\dag \hat{f}_i=1$ at each site, where $\bs \sigma=(\sigma_x,\sigma_y,\sigma_z)$.
The two and three spin interaction terms can be rewritten as the following by using the spinon hopping operator $\hat\chi_{ij} = \hat\chi_{ji}^\dag = \hat{f}_i^\dag \hat{f}_j$,
\begin{eqnarray}
\bold S_i\cdot \bold S_j &=& -{1\over2}\hat\chi_{ij} \hat\chi_{ji},\\
\mathbf{S}_{i}\cdot(\mathbf{S}_{j}\times\mathbf{S}_{k})&=& {1\over24i} \{ [\hat\chi_{ij}\hat\chi_{jk}\hat\chi_{ki} +\hat\chi_{jk}\hat\chi_{ij}\hat\chi_{ki} \nonumber\\
&&+ {\rm cyclic}(ijk)] -{\rm H.c.}\}.
\label{spinchiral}
\end{eqnarray}
In general the spinon hopping term is complex, and the spin chirality term can give rise to a phase $e^{i\phi_\Delta}$, with the flux $\phi_\Delta={\rm Arg}(\langle\hat\chi_{ik}\rangle\langle\hat\chi_{kj}\rangle\langle\hat\chi_{ji}\rangle)$ experienced by spinons after hopping through a closed minimum triangular loop in anticlockwise direction $i\rightarrow j\rightarrow k\rightarrow i$. When $\phi_\Delta=\frac{\pi}{2}$ the spinons experience a uniform U($1$) gauge field, with the magnetic flux through each triangular being $\frac{\pi}{2}$ and through each plaquette being $\pi$~\cite{Wen1989}.

The mean-field parameters about hopping can be introduced to decouple the spin interactions
(see Eqs.(\ref{hoppara})):
\begin{eqnarray*}
\chi_1
&=&
+\langle \hat\chi_{ii+1_x}\rangle^*|_{i_x={\rm odd},i_y={\rm odd}}
\\
&=&
-\langle \hat\chi_{ii+1_x}\rangle|_{i_x={\rm even}
,i_y={\rm odd}}
\\
&=&
+\langle \hat\chi_{ii+1_x}\rangle|_{i_x={\rm odd},i_y={\rm even}}
\\
&=&
-\langle \hat\chi_{ii+1_x}\rangle^*|_{i_x={\rm even}
,i_y={\rm even}}
\\
&=&
+\langle \hat\chi_{ii+1_y}\rangle|_{i_x={\rm odd},i_y={\rm odd}}
\\
&=&
-\langle \hat\chi_{ii+1_y}\rangle^*|_{i_x={\rm odd},i_y={\rm even}}\\
&=&
+\langle \hat\chi_{ii+1_y}\rangle^*|_{i_x={\rm even},i_y={\rm odd}}
\\
&=&
-\langle \hat\chi_{ii+1_y}\rangle|_{i_x={\rm even}
,i_y={\rm even}},
\\
\chi_2
&=&
+\langle \hat\chi_{ii+1_x+1_y}\rangle=\chi_2^*
\end{eqnarray*}
with $i_{x}=\frac{x_i}{a}$, $i_y=\frac{y_i}{a}$. Here $\chi_2$ is assumed to be real and the hopping phase is only carried by $\chi_1$ for the CSL phase.

Furthermore, since the system may contain symmetry breaking orders at some parameter region, we also need to introduce the following magnetic order parameters to describe the Neel order and stripe order, respectively
\begin{eqnarray}
M_n=(-1)^{i_x+i_y}\langle S^z_i\rangle, \ M_s = (-1)^{i_x}\langle S^z_i\rangle,
\end{eqnarray}

Now we can decouple the spin interactions by these trial mean-field parameters.
The first term in \eqref{effspinham3rdapp} is interaction of two spins on nearest neighbour sites
\begin{eqnarray}
\mathbf{S}_{i}\cdot \mathbf{S}_{j}&=&-\frac{1}{2}\langle\hat{\chi}_{ji}\rangle\hat{\chi}_{ij}+{\rm H.c.}
+\frac{1}{2}|\chi_1|^{2}\nonumber\\
&& +(-1)^{i_x+i_y} (M_n S^z_{j} - M_n S^z_i) + M_n^2,\nonumber\\
\end{eqnarray}
here $j=i+1_x $ or $j=i+1_y$.
And the second term in \eqref{effspinham3rdapp} is interaction of two spins on next nearest neighbour sites
\begin{eqnarray}
\mathbf{S}_{i}\cdot \mathbf{S}_{j}&=&-\frac{1}{2}\langle\hat{\chi}_{ji}\rangle\hat{\chi}_{ij}+{\rm H.c.}
+\frac{1}{2}|\chi_2|^{2}\nonumber\\
&& +(-1)^{i_x+i_y} (M_n S^z_{j} +M_n S^z_i) - M_n^2\nonumber\\
&& + (-1)^{i_x}(M_s S^z_{j}-M_s S^z_i) + M_s^2,
\end{eqnarray}
here $j=i+1_x +1_y$.

Both the Neel order $M_n$ and the stripe order $M_s$ are collinear, so they don't appear in decoupling
the spin chirality interaction term $\mathbf{S}_{i}\cdot(\mathbf{S}_{j}\times\mathbf{S}_{k})$
\begin{eqnarray}
&&\mathbf{S}_{i}\cdot(\mathbf{S}_{j}\times\mathbf{S}_{k})\nonumber\\
&=& \frac{1}{12i}[3\langle \hat\chi_{ij}\hat\chi_{jk}\rangle\hat\chi_{ki} + {\rm cyclic}(ijk) - {\rm H.c.}]\nonumber\\&&-{1\over12i}(6\langle\hat\chi_{ij}\rangle\langle\hat\chi_{jk}\rangle\langle\hat\chi_{ki}\rangle-{\rm H.c.})\nonumber\\
&=& {1\over12i}[3{|\chi_1^2\chi_2|e^{-i\phi_\Delta}\over\langle\hat\chi_{ki}\rangle}\hat\chi_{ki} + {\rm cyclic}(ijk) - {\rm H.c.}]
\nonumber\\
&&-{1\over12i}(6|\chi_1^2\chi_2|e^{-i\phi_\Delta}-{\rm H.c.})\label{3spinchiral}
\end{eqnarray}
with $\chi_{1}=|\chi_{1}|e^{i\frac{\pi}{4}}$. When we determine the matrix elements of the above three spin interaction term, we should keep in mind that every square plaquette contains four minimum triangular loops and every nearest neighbour bond is shared by four minimum triangular loops, but every next nearest neighbour bond is shared by only two minimum triangular loops.

When the spinons hop through a square plaquette in the direction and opposite direction of Fig.\ref{fourspin1}(a) of the main text, it will lead to the fourth part in \eqref{effspinham3rdapp}. It contains only nearest neighbour hopping.
$\phi_{ijkl}$ is the phase acquired by hopping through a closed square plaquette in anticlockwise direction $i\rightarrow j\rightarrow k\rightarrow l\rightarrow i$. It can be verified that $\phi_{ijkl}=\pi$ when $\phi_0=\frac{\pi}{4}$. The four spin interaction terms can be rewritten by the spinon hopping operator:
\begin{widetext}
\begin{eqnarray}
&&-\{80[(\mathbf{S}_{i}\cdot \mathbf{S}_{j})(\mathbf{S}_{k}\cdot\mathbf{S}_{l})+(\mathbf{S}_{i}\cdot \mathbf{S}_{l})(\mathbf{S}_{j}\cdot \mathbf{S}_{k})-(\mathbf{S}_{i}\cdot \mathbf{S}_{k})(\mathbf{S}_{j}\cdot \mathbf{S}_{l})] \nonumber\\
 && -4(\mathbf{S}_{i}\cdot \mathbf{S}_{j}+\mathbf{S}_{k}\cdot \mathbf{S}_{l}
      +\mathbf{S}_{j}\cdot \mathbf{S}_{k}+\mathbf{S}_{i}\cdot \mathbf{S}_{l}
      +\mathbf{S}_{i}\cdot \mathbf{S}_{k}+\mathbf{S}_{j}\cdot \mathbf{S}_{l})\}\nonumber\\
 &&=\hat{\chi}_{ij}\hat{\chi}_{jk}\hat{\chi}_{kl}\hat{\chi}_{li}+
 \hat{\chi}_{jk}\hat{\chi}_{kl}\hat{\chi}_{ij}\hat{\chi}_{li}+
 \hat{\chi}_{jk}\hat{\chi}_{ij}\hat{\chi}_{kl}\hat{\chi}_{li}+
 \hat{\chi}_{kl}\hat{\chi}_{jk}\hat{\chi}_{ij}\hat{\chi}_{li}+
 \frac{1}{2}(\hat{\chi}_{ij}\hat{\chi}_{kl}\hat{\chi}_{jk}\hat{\chi}_{li}+
 \hat{\chi}_{kl}\hat{\chi}_{ij}\hat{\chi}_{jk}\hat{\chi}_{li})+ {\rm cyclic}(ijkl)+{\rm H.c.},
\nonumber \\
\end{eqnarray}
\end{widetext}
here $j=i+1_x, k=i+1_x+1_y, l=i+1_y$. The four spin interaction terms can be decoupled by trial mean-field parameters as
\begin{widetext}
\begin{eqnarray}
&&-\{80[(\mathbf{S}_{i}\cdot \mathbf{S}_{j})(\mathbf{S}_{k}\cdot\mathbf{S}_{l})+(\mathbf{S}_{i}\cdot \mathbf{S}_{l})(\mathbf{S}_{j}\cdot \mathbf{S}_{k})-(\mathbf{S}_{i}\cdot \mathbf{S}_{k})(\mathbf{S}_{j}\cdot \mathbf{S}_{l})] \nonumber\\
 && -4(\mathbf{S}_{i}\cdot \mathbf{S}_{j}+\mathbf{S}_{k}\cdot \mathbf{S}_{l}
      +\mathbf{S}_{j}\cdot \mathbf{S}_{k}+\mathbf{S}_{i}\cdot \mathbf{S}_{l}
      +\mathbf{S}_{i}\cdot \mathbf{S}_{k}+\mathbf{S}_{j}\cdot \mathbf{S}_{l})\}\nonumber\\
 &=&5[4\langle\hat{\chi}_{ij}\rangle\langle\hat{\chi}_{jk}\rangle\langle\hat{\chi}_{kl}\rangle\hat{\chi}_{li}+
{\rm cyclic}(ijkl)-12\langle\hat{\chi}_{ij}\rangle\langle\hat{\chi}_{jk}\rangle\langle\hat{\chi}_{kl}\rangle
\langle\hat{\chi}_{li}\rangle+{\rm H.c.}]\nonumber\\
 &&-\{80[(\mathbf{S}_{i}\cdot \mathbf{S}_{j})(\mathbf{S}_{k}\cdot\mathbf{S}_{l})-(\mathbf{S}_{i}\times\mathbf{S}_{j})\cdot (\mathbf{S}_{k}
\times\mathbf{S}_{l})] \nonumber\\
 && -4(\mathbf{S}_{i}\cdot \mathbf{S}_{j}+\mathbf{S}_{k}\cdot \mathbf{S}_{l}
      +\mathbf{S}_{j}\cdot \mathbf{S}_{k}+\mathbf{S}_{i}\cdot \mathbf{S}_{l}
      +\mathbf{S}_{i}\cdot \mathbf{S}_{k}+\mathbf{S}_{j}\cdot \mathbf{S}_{l})\}\nonumber\\
 &=&5[\frac{4|\chi_1|^{4}e^{-i\pi}}{\langle\hat{\chi}_{li}\rangle}\hat{\chi}_{li}+
{\rm cyclic}(ijkl)-12|\chi_1|^{4}e^{-i\pi}+{\rm H.c.}]\nonumber\\
 &&-80[(-1)^{i_x+i_y}(M_{n}^{3}S^{z}_{i}-M_{n}^{3}S^{z}_{j}+M_{n}^{3}S^{z}_{k}-M_{n}^{3}S^{z}_{l})-3M_n^4]\nonumber\\
 &&-4(-1)^{i_x+i_y}(M_{n}S^{z}_{i}-M_{n}S^{z}_{j}+M_{n}S^{z}_{k}-M_{n}S^{z}_{l})+8M_n^2 .
\nonumber \\
\end{eqnarray}
\end{widetext}
When the spinons hop through a square plaquette in the direction and opposite direction of
Fig.\ref{fourspin1}(b) of the main text, it will lead to the fifth part in \eqref{effspinham3rdapp}.
It contains only next nearest neighbour hopping. The four spin interaction terms can also be rewritten by the spinon hopping operator:
\begin{widetext}
\begin{eqnarray}
&&80[(\mathbf{S}_{i}\cdot \mathbf{S}_{j})(\mathbf{S}_{k}\cdot\mathbf{S}_{l})+(\mathbf{S}_{i}\cdot \mathbf{S}_{l})(\mathbf{S}_{j}\cdot \mathbf{S}_{k})-(\mathbf{S}_{i}\cdot \mathbf{S}_{k})(\mathbf{S}_{j}\cdot \mathbf{S}_{l})] \nonumber\\
 && -4(\mathbf{S}_{i}\cdot \mathbf{S}_{j}+\mathbf{S}_{k}\cdot \mathbf{S}_{l}
      +\mathbf{S}_{j}\cdot \mathbf{S}_{k}+\mathbf{S}_{i}\cdot \mathbf{S}_{l}
      +\mathbf{S}_{i}\cdot \mathbf{S}_{k}+\mathbf{S}_{j}\cdot \mathbf{S}_{l})\nonumber\\
 &&=-\hat{\chi}_{ij}\hat{\chi}_{jk}\hat{\chi}_{kl}\hat{\chi}_{li}-
 \hat{\chi}_{jk}\hat{\chi}_{kl}\hat{\chi}_{ij}\hat{\chi}_{li}-
 \hat{\chi}_{jk}\hat{\chi}_{ij}\hat{\chi}_{kl}\hat{\chi}_{li}-
 \hat{\chi}_{kl}\hat{\chi}_{jk}\hat{\chi}_{ij}\hat{\chi}_{li}-
 \frac{1}{2}(\hat{\chi}_{ij}\hat{\chi}_{kl}\hat{\chi}_{jk}\hat{\chi}_{li}+
 \hat{\chi}_{kl}\hat{\chi}_{ij}\hat{\chi}_{jk}\hat{\chi}_{li})+ {\rm cyclic}(ijkl)+{\rm H.c.},
\nonumber \\
\end{eqnarray}
\end{widetext}
here $j=i+1_x+1_y, k=i+2_y, l=i-1_x+1_y$. The four spin interaction terms can be decoupled by trial mean-field parameters as
\begin{widetext}
\begin{eqnarray}
&&80[(\mathbf{S}_{i}\cdot \mathbf{S}_{j})(\mathbf{S}_{k}\cdot\mathbf{S}_{l})+(\mathbf{S}_{i}\cdot \mathbf{S}_{l})(\mathbf{S}_{j}\cdot \mathbf{S}_{k})-(\mathbf{S}_{i}\cdot \mathbf{S}_{k})(\mathbf{S}_{j}\cdot \mathbf{S}_{l})] \nonumber\\
 && -4(\mathbf{S}_{i}\cdot \mathbf{S}_{j}+\mathbf{S}_{k}\cdot \mathbf{S}_{l}
      +\mathbf{S}_{j}\cdot \mathbf{S}_{k}+\mathbf{S}_{i}\cdot \mathbf{S}_{l}
      +\mathbf{S}_{i}\cdot \mathbf{S}_{k}+\mathbf{S}_{j}\cdot \mathbf{S}_{l})\nonumber\\
 &=&-5[4\langle\hat{\chi}_{ij}\rangle\langle\hat{\chi}_{jk}\rangle\langle\hat{\chi}_{kl}\rangle\hat{\chi}_{li}+
{\rm cyclic}(ijkl)\nonumber\\
 &&-12\langle\hat{\chi}_{ij}\rangle\langle\hat{\chi}_{jk}\rangle\langle\hat{\chi}_{kl}\rangle
\langle\hat{\chi}_{li}\rangle+{\rm H.c.}]\nonumber\\
 &&+80[(\mathbf{S}_{i}\cdot \mathbf{S}_{j})(\mathbf{S}_{k}\cdot\mathbf{S}_{l})-(\mathbf{S}_{i}\times\mathbf{S}_{j})\cdot (\mathbf{S}_{k}
\times\mathbf{S}_{l})] \nonumber\\
 && -4(\mathbf{S}_{i}\cdot \mathbf{S}_{j}+\mathbf{S}_{k}\cdot \mathbf{S}_{l}
      +\mathbf{S}_{j}\cdot \mathbf{S}_{k}+\mathbf{S}_{i}\cdot \mathbf{S}_{l}
      +\mathbf{S}_{i}\cdot \mathbf{S}_{k}+\mathbf{S}_{j}\cdot \mathbf{S}_{l})\nonumber\\
 &=&-5[4\chi_2^{3}\hat{\chi}_{li}+
{\rm cyclic}(ijkl)-12\chi_2^{4}+{\rm H.c.}]\nonumber\\
 &&+80[(-1)^{i_x+i_y}(M_{n}^{3}S^{z}_{i}+M_{n}^{3}S^{z}_{j}+M_{n}^{3}S^{z}_{k}+M_{n}^{3}S^{z}_{l})-3M_n^4]\nonumber\\
 &&-12(-1)^{i_x+i_y}(M_{n}S^{z}_{i}+M_{n}S^{z}_{j}+M_{n}S^{z}_{k}+M_{n}S^{z}_{l})+24M_n^2\nonumber\\
 &&+80[(-1)^{i_x}(M_{s}^{3}S^{z}_{i}-M_{s}^{3}S^{z}_{j}+M_{s}^{3}S^{z}_{k}-M_{s}^{3}S^{z}_{l})-3M_s^4]\nonumber\\
 &&+4(-1)^{i_x}(M_{s}S^{z}_{i}-M_{s}S^{z}_{j}+M_{s}S^{z}_{k}-M_{s}S^{z}_{l})-8M_s^2  .
\nonumber \\
\end{eqnarray}
\end{widetext}
Note that both the Neel order $M_n$ and the stripe order $M_s$ don't appear in decoupling
the four spin interaction term $(\mathbf{S}_{i}\times\mathbf{S}_{j})\cdot (\mathbf{S}_{k}
\times\mathbf{S}_{l})$ because they are collinear.

The terms about hopping events $i\leftrightarrow j, k\leftrightarrow l$ appear in the last two parts of \eqref{effham}: $T_{-}T_{+}T_{-}T_{+}-\frac{1}{2}T_{-}T_{-}T_{+}T_{+}$, but they are canceled by each other.

The first two terms in the last line of \eqref{effspinham3rdapp} are about interaction of
two spins located at two neighbour sites in Fig.\ref{fourspin2} of the main text. Each spin is interacted twice.

Using spinon hopping operators to rewrite the spin-interaction term:
\begin{eqnarray}
& &-16\mathbf{S}_{i}\cdot \mathbf{S}_{j}\nonumber\\
&=&2\sum_{\sigma}(\hat{\chi}_{ij}\hat{f}^{\dagger}_{j\sigma}\hat{f}_{i\sigma}
\hat{f}^{\dagger}_{i\sigma}\hat{f}_{j\sigma}\hat{\chi}_{ji}+
\hat{\chi}_{ji}\hat{f}^{\dagger}_{i\bar{\sigma}}\hat{f}_{j\bar{\sigma}}
\hat{f}^{\dagger}_{i\sigma}\hat{f}_{j\sigma}\hat{\chi}_{ji}).
\nonumber \\
\end{eqnarray}
Under mean-field approximation,
\begin{eqnarray}
&&\sum_{\sigma}\hat{\chi}_{ij}\hat{f}^{\dagger}_{j\sigma}\hat{f}_{i\sigma}
\hat{f}^{\dagger}_{i\sigma}\hat{f}_{j\sigma}\hat{\chi}_{ji}
\nonumber\\&=&
\sum_{\sigma}[\langle\hat{\chi}_{ij}\rangle\langle\hat{f}^{\dagger}_{j\sigma}\hat{f}_{i\sigma}\rangle
\langle\hat{f}^{\dagger}_{i\sigma}\hat{f}_{j\sigma}\rangle\hat{\chi}_{ji}
\nonumber\\
&&+\langle\hat{\chi}_{ij}\rangle\langle\hat{f}^{\dagger}_{j\sigma}\hat{f}_{i\sigma}\rangle
\hat{f}^{\dagger}_{i\sigma}\hat{f}_{j\sigma}\langle\hat{\chi}_{ji}\rangle
\nonumber\\
&&+\langle\hat{\chi}_{ij}\rangle\hat{f}^{\dagger}_{j\sigma}\hat{f}_{i\sigma}
\langle\hat{f}^{\dagger}_{i\sigma}\hat{f}_{j\sigma}\rangle\langle\hat{\chi}_{ji}\rangle
\nonumber\\
&&+\hat{\chi}_{ij}\langle\hat{f}^{\dagger}_{j\sigma}\hat{f}_{i\sigma}\rangle
\langle\hat{f}^{\dagger}_{i\sigma}\hat{f}_{j\sigma}\rangle\langle\hat{\chi}_{ji}\rangle
\nonumber\\
&&-3\langle\hat{\chi}_{ij}\rangle\langle\hat{f}^{\dagger}_{j\sigma}\hat{f}_{i\sigma}\rangle
\langle\hat{f}^{\dagger}_{i\sigma}\hat{f}_{j\sigma}\rangle\langle\hat{\chi}_{ji}\rangle]
\nonumber\\
&=&|\langle\hat{\chi}_{ij}\rangle|^{2}(\langle\hat{\chi}_{ij}\rangle\hat{\chi}_{ji}
+\langle\hat{\chi}_{ji}\rangle\hat{\chi}_{ij})-\frac{3}{2}|\langle\hat{\chi}_{ij}\rangle|^{4},
\nonumber \\
&&
\sum_{\sigma}\hat{\chi}_{ji}\hat{f}^{\dagger}_{i\bar{\sigma}}\hat{f}_{j\bar{\sigma}}
\hat{f}^{\dagger}_{i\sigma}\hat{f}_{j\sigma}\hat{\chi}_{ji}
\nonumber\\
&=&|\langle\hat{\chi}_{ij}\rangle|^{2}(\langle\hat{\chi}_{ij}\rangle\hat{\chi}_{ji}
+\langle\hat{\chi}_{ji}\rangle\hat{\chi}_{ij})-\frac{3}{2}|\langle\hat{\chi}_{ij}\rangle|^{4},
\nonumber \\
\end{eqnarray}
where we use $\langle\hat{f}^{\dagger}_{i\uparrow}\hat{f}_{j\uparrow}\rangle
=\langle\hat{f}^{\dagger}_{i\downarrow}\hat{f}_{j\downarrow}\rangle=\frac{1}{2}\langle\hat{\chi}_{ij}\rangle$.

We can decouple the two terms by trial mean-field parameters
\begin{eqnarray}
-16\mathbf{S}_{i}\cdot \mathbf{S}_{j}&=&4|\chi_{1}|^{2}(\langle\hat{\chi}_{ij}\rangle\hat{\chi}_{ji}
+\langle\hat{\chi}_{ji}\rangle\hat{\chi}_{ij})-6|\chi_{1}|^{4}\nonumber\\
&& -16[(-1)^{i_x+i_y} (M_n S^z_{j} - M_n S^z_i) + M_n^2],
\nonumber \\
\end{eqnarray}
here $j=i+1_x $ or $j=i+1_y$,
\begin{eqnarray}
-16\mathbf{S}_{i}\cdot \mathbf{S}_{j}&=&4\chi_{2}^{2}(\langle\hat{\chi}_{ij}\rangle\hat{\chi}_{ji}
+\langle\hat{\chi}_{ji}\rangle\hat{\chi}_{ij})-6\chi_{2}^{4}\nonumber\\
&& -16[(-1)^{i_x+i_y} (M_n S^z_{j} +M_n S^z_i) - M_n^2]\nonumber\\
&& -16[ (-1)^{i_x}(M_s S^z_{j}-M_s S^z_i) + M_s^2],
\nonumber \\
\end{eqnarray}
here $j=i+1_x +1_y$.

The three spins located at the three sites of minimum triangular loop also give rise to four spin interactions,
in which one spin is interacted twice while the other two spins are interacted only once. They are illustrated in Fig.\ref{fourspin3} of the main text and correspond to
the last two terms in the last line of \eqref{effspinham3rdapp}.
The two terms can be obtained by using spinon hopping operators to rewrite the spin interaction term:
\begin{eqnarray}\label{4spin4}
&&4\mathbf{S}_{j}\cdot \mathbf{S}_{l}+{\rm cyclic}(ijl)\nonumber\\
&=&\{\sum_{\sigma}(\hat{\chi}_{il}\hat{f}^{\dagger}_{l\sigma}\hat{f}_{i\sigma}
\hat{f}^{\dagger}_{i\sigma}\hat{f}_{j\sigma}\hat{\chi}_{ji}
+\hat{\chi}_{ij}\hat{f}^{\dagger}_{j\sigma}\hat{f}_{i\sigma}
\hat{f}^{\dagger}_{l\bar{\sigma}}\hat{f}_{i\bar{\sigma}}\hat{\chi}_{il}\nonumber\\
&&+\hat{\chi}_{li}\hat{f}^{\dagger}_{i\bar{\sigma}}\hat{f}_{l\bar{\sigma}}
\hat{f}^{\dagger}_{i\sigma}\hat{f}_{j\sigma}\hat{\chi}_{ji}
+\hat{\chi}_{ji}\hat{f}^{\dagger}_{i\sigma}\hat{f}_{j\sigma}
\hat{f}^{\dagger}_{l\sigma}\hat{f}_{i\sigma}\hat{\chi}_{il}+j\leftrightarrow l)\nonumber\\
&&-[(\sum_{\sigma}\hat{\chi}_{li}\hat{f}^{\dagger}_{i\sigma}\hat{f}_{j\sigma}
\hat{f}^{\dagger}_{i\bar{\sigma}}\hat{f}_{l\bar{\sigma}}\hat{\chi}_{ji}+j\leftrightarrow l)+{\rm H.c.}]\nonumber\\
&&-\sum_{\sigma}(\hat{\chi}_{ij}\hat{f}^{\dagger}_{l\sigma}\hat{f}_{i\sigma}
\hat{f}^{\dagger}_{i\sigma}\hat{f}_{l\sigma}\hat{\chi}_{ji}
+\hat{\chi}_{ji}\hat{f}^{\dagger}_{i\sigma}\hat{f}_{l\sigma}
\hat{f}^{\dagger}_{l\sigma}\hat{f}_{i\sigma}\hat{\chi}_{ij}\nonumber\\
&&+j\leftrightarrow l)\}+{\rm cyclic}(ijl),
\end{eqnarray}
where the term $4\mathbf{S}_{j}\cdot \mathbf{S}_{l}$ corresponds to spin interaction illustrated in
Fig.\ref{fourspin3} (a), which contains only nearest neighbour hopping,
while the terms $4\mathbf{S}_{i}\cdot \mathbf{S}_{l}$ and $4\mathbf{S}_{i}\cdot \mathbf{S}_{j}$ in ${\rm cyclic}(ijl)$ correspond to spin interactions illustrated in Fig.\ref{fourspin3} (b) and (c),
which contain not only nearest neighbour hopping, but next nearest neighbour hopping. Also note that every nearest neighbour bond is shared by four minimum triangular loops, but every next nearest neighbour bond is shared by only two minimum triangular loops, we can thus obtain the last two terms of \eqref{effspinham3rdapp}. When using trial mean-field parameters to decouple the spin interaction term \eqref{4spin4}, we find

\begin{eqnarray}
&&\sum_{\sigma}(\hat{\chi}_{il}\hat{f}^{\dagger}_{l\sigma}\hat{f}_{i\sigma}
\hat{f}^{\dagger}_{i\sigma}\hat{f}_{j\sigma}\hat{\chi}_{ji}
+\hat{\chi}_{ij}\hat{f}^{\dagger}_{j\sigma}\hat{f}_{i\sigma}
\hat{f}^{\dagger}_{l\bar{\sigma}}\hat{f}_{i\bar{\sigma}}\hat{\chi}_{il}\nonumber\\
&&+\hat{\chi}_{li}\hat{f}^{\dagger}_{i\bar{\sigma}}\hat{f}_{l\bar{\sigma}}
\hat{f}^{\dagger}_{i\sigma}\hat{f}_{j\sigma}\hat{\chi}_{ji}
+\hat{\chi}_{ji}\hat{f}^{\dagger}_{i\sigma}\hat{f}_{j\sigma}
\hat{f}^{\dagger}_{l\sigma}\hat{f}_{i\sigma}\hat{\chi}_{il}+j\leftrightarrow l)\nonumber\\
&&-[(\sum_{\sigma}\hat{\chi}_{li}\hat{f}^{\dagger}_{i\sigma}\hat{f}_{j\sigma}
\hat{f}^{\dagger}_{i\bar{\sigma}}\hat{f}_{l\bar{\sigma}}\hat{\chi}_{ji}+j\leftrightarrow l)+{\rm H.c.}]\nonumber\\
&&-\sum_{\sigma}(\hat{\chi}_{ij}\hat{f}^{\dagger}_{l\sigma}\hat{f}_{i\sigma}
\hat{f}^{\dagger}_{i\sigma}\hat{f}_{l\sigma}\hat{\chi}_{ji}
+\hat{\chi}_{ji}\hat{f}^{\dagger}_{i\sigma}\hat{f}_{l\sigma}
\hat{f}^{\dagger}_{l\sigma}\hat{f}_{i\sigma}\hat{\chi}_{ij}\nonumber\\
&&+j\leftrightarrow l)\nonumber\\
&=&[4|\langle\hat{\chi}_{ij}\rangle|^{2}(\langle\hat{\chi}_{il}\rangle\hat{\chi}_{li}
+\langle\hat{\chi}_{li}\rangle\hat{\chi}_{il})+j\leftrightarrow l\nonumber\\
&&-12|\langle\hat{\chi}_{ij}\rangle|^{2}|\langle\hat{\chi}_{il}\rangle|^{2}]\nonumber\\
&&-[4|\langle\hat{\chi}_{ij}\rangle|^{2}(\langle\hat{\chi}_{il}\rangle\hat{\chi}_{li}
+\langle\hat{\chi}_{li}\rangle\hat{\chi}_{il})+j\leftrightarrow l\nonumber\\
&&-12|\langle\hat{\chi}_{ij}\rangle|^{2}|\langle\hat{\chi}_{il}\rangle|^{2}]\nonumber\\
&=&0.
\end{eqnarray}
Therefore the mean-field approximation of hopping terms is equal to zero. We can only use magnetic order parameters to decouple the last two terms of \eqref{effspinham3rdapp}:
\begin{eqnarray}
16\mathbf{S}_{i}\cdot \mathbf{S}_{j}&=&16[(-1)^{i_x+i_y} (M_n S^z_{j} - M_n S^z_i) + M_n^2],
\nonumber \\
\end{eqnarray}
here $j=i+1_x $ or $j=i+1_y$,
\begin{eqnarray}
8\mathbf{S}_{i}\cdot \mathbf{S}_{j}&=&8[(-1)^{i_x+i_y} (M_n S^z_{j} +M_n S^z_i) - M_n^2],
\nonumber \\
\end{eqnarray}
here $j=i+1_x +1_y$.


\normalem


\end{document}